\documentclass[journal,comsoc]{IEEEtran}

\usepackage[T1]{fontenc}
\ifCLASSINFOpdf
\else 
\fi
\usepackage{amsmath}
\usepackage[cmintegrals]{newtxmath}
\usepackage{graphicx}
\usepackage{multirow}
\usepackage{subfigure}
\usepackage[noadjust]{cite}
\usepackage{enumerate}
\usepackage{diagbox}
\usepackage{makecell}
\usepackage{blindtext}
\usepackage{enumitem}
\usepackage{multirow}
\usepackage{footnote}
\makesavenoteenv{tabular}
\makesavenoteenv{table}
\usepackage[hang,flushmargin]{footmisc}

\usepackage[linesnumbered,ruled,vlined]{algorithm2e}

\SetCommentSty{mycommfont}

\usepackage[dvipsnames,table,xcdraw]{xcolor}
\usepackage{hhline} 
\usepackage{highlight}

\let\oldnl\nl
\newcommand{\nonl}{\renewcommand{\nl}{\let\nl\oldnl}}
\renewcommand{\opacity}{100}

\usepackage{lipsum}
\usepackage{fancyhdr}

\begin{document}
	
	\title{Multi-objective Scheduling of Electric Vehicle Charging/Discharging with Time of Use Tariff
		\thanks{This work is supported by Australian Research Council through project no LP180101309.}}
	
	\author{Hui~Song,
		Chen~Liu,
		Mahdi~Jalili,
		Xinghuo~Yu,
		Peter~McTaggart 
		\thanks{Hui Song, Chen Liu, Mahdi Jalili, Xinghuo Yu, School of Engineering, RMIT University, Australia, Melbourne, hui.song@rmit.edu.au, chen.liu3@rmit.edu.au, mahdi.jalili@rmit.edu.au, xinghuo.yu@rmit.edu.au.}
		\thanks{Peter~McTaggart, CitiPower and Powercor, Australia, Melbourne, pmctaggart@powercor.com.au.}}

	\markboth{IEEE Transactions on Smart Grid}%
	{Shell \MakeLowercase{\textit{et al.}}: Bare Demo of IEEEtran.cls for IEEE Communications Society Journals}
	
	\maketitle

	\begin{abstract}
		
		The increased uptake of electric vehicles (EVs) leads to increased demand for electricity, and sometimes pressure on power grids. Uncoordinated charging of EVs may result in stress on distribution networks, and often some form of optimization is required in the charging process. Optimal coordinated charging is a multi-objective optimization problem (MOOP) in nature, with objective functions such as minimum price charging and minimum disruptions to the grid. In this manuscript, we propose a general multi-objective EV charging/discharging schedule (MOEVCS) framework, where the time of use (TOU) tariff is designed according to the load request at each time stamp. To obtain the optimal scheduling scheme and balance the competing benefits from different stakeholders, such as EV owners, EV charging stations (EVCS), and the grid operator, we design three competing objective functions including EV owner cost, EVCS profit, and the network impact. Moreover, we create four application scenarios with different charging request distributions over the investigated periods. Due to different types of decision variables in this MOOP, we develop a constraint mixed-variable multi-objective evolutionary algorithm (MVMOEA) to implement the proposed MOEVCS framework. Our results demonstrate the effectiveness of MOEVCS in making a balance between three competing objectives.  
		
	\end{abstract}
	
	
	\IEEEpeerreviewmaketitle
	

	\section{Introduction}
	
	As a promising component of sustainable and eco-friendly transportation system~\cite{das2020multi, sundstrom2011flexible}, electric vehicles (EVs) have received much attention in many countries~\cite{pagany2019review}. According to the report of International Energy Agency\footnote{https://webstore.iea.org/global-ev-outlook-2019}, the global EV fleet exceeded 5.1 million in 2018, and the stock may reach 250 million by 2030. The additional load on the electrical power systems due to the growth of EV adoption changes the electricity consumption profiles and leads to challenges, such as energy loss, overload, load unbalances, load peaks, harmonic distortion voltage instability, negative impact on the grid network. More EVs joining to the electricity networks challenges demand management in the distribution network, especially during the off-peak hours~\cite{leou2015optimal, ortega2014optimal}. For example, many EVs being coincidently charged on the same low voltage network may lead to severe impacts to network and other energy consumers on that network. Another challenge arising from EV charging is the increased demand for the public charging stations (CSs), CS placement and sizing~\cite{parastvand2020graph, hashemian2020pev}. Therefore, how to effectively charge EV while considering the benefits for both EV users and CSs and the impact on the distribution network is becoming a crucial concern for network operators~\cite{liu2019coordinated}. Coordinated charging, which aims to determine the charging schedule for each EV and considers the overall impact caused by large-scale charging requests during a specific period, provides a promising approach to address this urgent issue \cite{xu2015hierarchical, jian2017high, zhou2020admm, zhou2020coordinated}.     
	
	Some research efforts have been made toward developing the optimal coordinated charging/discharging strategies for EVs by incorporating the benefits of different stakeholders. To minimize the power losses and maximize the main grid load factor, coordinated charging was proposed in~\cite{clement2009impact} to obtain the optimal charging profile of the plug-in hybrid electric vehicles (PHEVs). EV coordination and vehicle-to-grid (V2G) resource optimization of the EV aggregator was formulated in~\cite{nizami2020coordinated}, where a mixed-integer programming-based optimization model was applied to minimize the cost of EV owners using a real-time tariff by considering the local grid constraints to reduce the negative impact of charging. Considering the dependency of the EV charging stations (EVCSs), the charging options at the available stations, and the charging amount,~\cite{liu2019coordinated} formulated the objective function by aggregating three objectives including the time, expense, and State of Charge (SoC) gap. A mixed-variable differentiate evolution (MVDE) was proposed to obtain the dynamic charging scheduling for all EVs at a group of CSs on a transportation network~\cite{liu2019coordinated}. Most of the existing works modelled the coordinated charging as a single-objective optimization (SOO) problem with constraints like distribution network impact, gird-level impact, and CS load capacity while ignoring the competing benefits between different stakeholders and the network impact. However, in real-world applications, the EV charging schedule is modelled as a multi-objective optimization (MOO) problem with a number of competing objective functions.    
	
	Some works have framed the charge scheduling as an MOO problem.~\cite{kang2017optimal} used load stabilizing function to reduce the fluctuation of power grid and economic loss as two competing objective functions to solve the optimal load scheduling problem.~\cite{shukla2019multi} formulated the power loss, voltage deviation of the distribution system, and EV flow served by the fast charging station as the competing objectives to obtain the final planning scheme, where an MOO approach was applied to obtain the non-dominated Pareto front (PF) of the solutions. Some of the existing works are only based on the simulated data~\cite{faddel2018automated}, and more attention should be paid to collect the real-world data. Moreover, most of the prior studies only focus on the investigated scenarios without comparing to some general baselines. With the aim to flatten the consumption peaks, dynamic time of use (TOU) tariff structure~\cite{zhang2018optimal, nizami2020coordinated} have been frequently used by many retailers. Under TOU tariff, many EV owners choose to charge the EVs during off-peak hours, and discharge (if vehicle to grid connectivity is enabled) during the peak hours. The benefits of EVCS and the cost of EV owners should be considered simultaneously to avoid the price spikes during some periods. 
	
	In this manuscript, we propose a general multi-objective EV charging/discharging schedule (MOEVCS) framework for EVCSs, where dynamic TOU tariff is designed according to the load demand during different time periods. Three competing objective functions are designed in MOEVCS. We design a mixed-variable multi-objective evolutionary algorithm (MVMOEA) based on the constraint non-dominated sorting genetic algorithm (NSGA-II)~\cite{deb2000fast} to solve this MOO problem. The main contributions are summarized as follows: 
	
	\begin{itemize}
		\item We propose a general MOEVCS framework, which can provide the optimal charging/discharging schedule for large number of EV owners. 
		\item We design three competing objective functions, i.e., to minimize the cost of the EV owners, maximize the benefits of the EVCS, and minimize the load variation, to balance the benefits between the EV owners and EVCS as well as decrease the negative impacts on the grid network. 
		\item We develop an MVMOEA to implement the proposed MOEVCS subjected to the designed competing objective functions, given that the discrete and continuous variables involved in this optimization problem.  
		\item We propose a TOU tariff that is changed with the load demand during each time period in case that many EV owners choose to charge during the off-peak hours. The TOU tariff may contribute to the network stability since it can flatten the load fluctuation significantly.  
		\item To verify the effectiveness of MOEVCS on balancing the benefits from different stakeholder by solving the MOO problem, we create four different problem sets for simulation and design five baselines for comparisons. 
		
	\end{itemize}

	The rest of this paper is organized as follows. The related work is discussed in Section~\ref{related_works}. Section~\ref{problem_modeling} presents the problem formulation, and Section~\ref{proposed_method} introduces the proposed framework and its implementation. Experimental results are reported and discussed in Section~\ref{experiments}. Section~\ref{conclusion} concludes the paper with some future work being mentioned.
	
	\section{Related Works}\label{related_works}
	
	 We mainly focus on the optimization of coordinated charging/discharging in the public charging stations. 
	
	\subsection{Applications of Optimization Methods in Coordinated Charging of EVs}
	The growing EV penetration may lead to pressure on the distribution networks, if the charging is uncoordinated. Often, some ways of optimization is applied to coordinate the charging while satisfying some objective functions. Some optimization techniques have been applied for EV coordinated charging problems. The EV fast charging problem was modelled as an optimization problem subjected to the coupled feeder capacity constraints in the distribution network, solved by a decentralized iterative algorithm based on the gradient projection methods~\cite{zhou2020admm}. A mixed-integer programming-based optimization model was used to minimize the cost of EV owners in~\cite{nizami2020coordinated} for EV coordination and V2G resource optimization. MVDE was proposed in~\cite{liu2019coordinated} to obtain the dynamic charging scheduling by aggregating three different objective functions, such as time cost, expense, and SoC as an SOO problem. Quadratic programming was applied to minimize the power losses and maximize the main grid load factor for the optimal charging profile of PHEVs~\cite{clement2009impact}. Linear programming was used to minimize EVs' energy cost based on the day-ahead price with constraints such as SoC limits, the EVs maximum power, the battery SoC after charging, and the distribution feeder subscribed power in the residential area~\cite{ayyadi2019optimal}. 
	
	Coordinated charging often involves competing benefits from different stakeholders and impacts on the distribution network and the power grid, but existing works, including those mentioned above, have mainly focused on SOO with the relevant constraints, ignoring benefit or impact on the competing aspects. Although~\cite{liu2019coordinated} considered three objectives, they were converted into an SOO problem by using a simple linear combination, i.e., no MOO technique was used. This fails to handle competing objectives and leads to another challenge that is to manually choose suitable weights to balance different objectives. MOO provides the potential solution to address the optimization problem involved more than one competing objective function, given that it can generate a set of trade-off solutions that are no better than any other one in the entire solution space. A weight aggregation (WA) strategy based MO particle swarm optimization (WA-MOPSO) was proposed in~\cite{kang2017optimal} to minimize two competing objectives, i.e., potential serious peak-to-valley difference and economic loss. MO grey wolf optimizer (MOGWO) was used in \cite{shukla2019multi} to achieve the planning scheme subjected to the power losses, voltage deviation of the distribution system, and the EV flow. Compared to SOO, MOO is a more promising technique in the coordinated charging problem due to the objectives involved. Since the complexity of coordinated charging is being improved with the growing EV penetration, more advanced MOO algorithms will be developed in the future.         
	
	\subsection{Optimization in Public Charging Stations}
	With the increasing development of EVs, the charging infrastructure has been set up at locations with high charging potential~\cite{sufyan2020charge}. Some existing works have investigated the challenges caused by public coordinated charging, which may involve charger option, transportation network, traffic flow, profit of charging stations, and impact on the power grid and distribution network. A MO collaborative planning strategy was proposed in~\cite{yao2014multi} to minimize the annual cost of investment and energy losses and maximize the traffic flow to provide better service by reducing the waiting time at the station. Considering the need for fast charging and reducing degradation in batteries and the distribution network, a hierarchical model was proposed to achieve an optimal strategy profile for EVs and verified on a 5-feeder and a 12-feeder test systems~\cite{zhou2020admm}. The charging schedule for each charging pile over each hour during one day was investigated subjected to two competing objective functions, i.e., fluctuation of the power grid and the revenue, where WA-MOPSO was proposed to address the formulated MOO problem~\cite{kang2017optimal}. 
	
	\cite{zhang2018optimal} focused on designing an optimal pricing scheme that can guide and coordinate the charging processes of EVs in the CS to minimize the service dropping rate of the CS. By optimizing the energy losses, voltage deviation, and the served traffic flow,~\cite{shukla2019multi} aimed to develop a spatial-temporal distribution of the vehicles on the road to find the nearest public CS with the available battery SoC.~\cite{chen2018coordinated} proposed an optimal real-time coordinated charging and discharging strategy for plug-in electric bus fast CSs (PEBFCSs) with energy storage system (ESS) to achieve the maximum economic benefits considering the lifespan of ESS, the capacity charge of PEBFCS, and the electricity price arbitrage.~\cite{wan2020game} targeted at building one model to coordinate the charging strategies of all PEVs such that the energy cost of the smart charging station was minimized without compromising a set of constraints for EVs and the CS, addressed by a distributed generalized Nash equilibrium-seeking algorithm based on the Newton fixed-point method. The prior works have mainly focused on the optimal charging pile scheduling, the optimal TOU tariff that can motive coordinated charging in the CS, and the most economic CS for EV users by evaluating the served EVs. 
	
	\subsection{Research Gaps}
	The studies related to coordinated charging are not limited to the references mentioned here since the development of EV coordinated charging is still ongoing, and there are still many research gaps to be filled. EVs are publicly used in some countries, some of which are based on simulated data or distribution network data, and collecting real-world data for research is still essential. Many prior works also suffer from lack of baselines to compare with. Some works only consider the benefit of EVCS while ignoring the cost of EV users. The coordinated charging problems always involve multiple competing objectives, and developing effective MOO algorithms to find the trade-off solution is largely missing in the open literature. Limited works have considered coordinated charging and discharging together for an EV. Building a model that can optimally choose to charge or discharge during each time slot is another challenge to be carefully studied. The benefit of EVCSs and coordinated charging impacts on the distribution network and power grid is influenced by not only a single EV, but optimal charging/discharging schedule over all potential EVs during the investigated time slots. These gaps motivate us to propose the MOEVCS framework subjected to three competing objectives including the EV users' cost, EVCS's profit, and the network impact, where TOU tariff is designed according to the load demand during different time periods. Considering the mixed variables involved, we develop an MVMOEA to implement MOEVCS. We also develop five baselines and four different application scenarios using a real-world dataset, to verify the effectiveness of MOEVCS.

	\section{Problem Formulation}\label{problem_modeling}
	
	The EV users' cost, the EVCS'S benefit, and the network impact over the optimal EV charging/discharging scenario are modelled, where the pricing mechanism in EV charging/discharging market is introduced.

	\subsection{EV User Charging Cost Modelling}\label{EV_user_modeling}
	
	EV user charging/discharging scenario has two requirements: i) each user would like the EV to be fully charged at the departure time; ii) the charging/discharging activities lead to the minimal cost. Assuming that the EV parking time is longer than the charging time, the EV has the opportunity to reduce the charging cost by considering dynamic electricity prices at different time slots, so that the EV user can select the lower price time slot rather than being fully charged in a short period of time. This is achievable provided that the EVCS has enough parking lot, especially in large public parking equipped with charging facilities. In addition, the cost can be further reduced by exporting from EV's battery to the EVCS or power grid. The EV user generates profits by selling electricity in peak time slots and performing charging activities during off-peak time slots, but the EV charging behaviour may have an impact on the prices. If all EV users charge EVs at off-peak price time and discharge at peak price period, the price will be increased at charging time and decreased at discharging time (i.e., peak and off-peak will change). In other words, EV users cannot reduce the charging cost if they make the decision to schedule the charging/discharging time independently. It is challenging to arrange the charging and discharging schedule for all EV users together in one EVCS to reduce the overall cost for all EV users. Another EV charging and discharging challenge is battery degradation caused by frequent charging and discharging activities. In some cases, the EV user might end up paying more than the discharging profit for a new battery. Thus, battery degradation should be considered in the EV charging/discharging activity.
	
	The EV user cost includes the charging cost, discharging benefit, and battery degradation cost. For an EV $e \in[1, N_\textit{user}]$ in the EVCS, it has a set of parking time $T’$. Let $x^c_{e,t’}$ and $x^d_{e,t’}$ be the binary variable to indicate charging and discharging state for $e$ in the parking time slot $t’$, respectively. $x^c_{e,t’}$/$x^d_{e,t’}$ equals to 1 and means that the EV is charging/discharging at $t’$, otherwise, $x^c_{e,t’}$/$x^d_{e,t’}$ equals to 0. Considering EV charging power $x^{c,s}_{e,t'}$, discharging power $x^{d,s}_{e,t'}$, electricity price $\lambda_{t’}$ at each $t’$, and battery degradation cost~\cite{ortega2014optimal}, the total cost for EV $e$ is minimized and can be modelled as follows:
	\begin{eqnarray}
		\min f_1(x^c,x^{c,s},x^d,x^{d,s}) = \sum\limits_{t' \in T'} [(x^c_{e,t'} x^{c,s}_{e,t'}+x^d_{e,t'} x^{d,s}_{e,t'})\lambda_{t'}\nonumber \\
		+ |\frac{k_e}{100}|\frac{(SoC_{e,t'-1}-SoC_{e,t'})_+}{B_e} C^B_e].\label{eq:EV_obj}
	\end{eqnarray}
	The first term of Eq.~\ref{eq:EV_obj} denotes the cost from EV charging and discharging activities. The second term represents battery degradation cost. For any EV at EVCS, there is only one state among three states including charging ($x^{c,s}_{e,t’} = 1, x^{d,s}_{e,t’} = 0$), discharging ($x^{c,s}_{e,t’} = 0, x^{d,s}_{e,t’} = -1$) and idle ($x^{c,s}_{e,t’} = x^{d,s}_{e,t’} = 0$). An EV can only have one state at any time slot. The $e$th EV battery state at $t’$ is constrained as follows:
	\begin{eqnarray}
		\ x^{c,s}_{e,t'} \in \{0, 1\},x^{d,s}_{e,t'} \in \{0, -1\},
		x^{c,s}_{e,t'}+|x^{d,s}_{e,t'}| \leq 1.\label{eq:EV_con2}
	\end{eqnarray}
	To satisfy the EV charging demand, the SoC should achieve the EV user's requirement, modelled as follows: 
	\begin{eqnarray}
		SoC^{\textit{arr}} + \phi \sum\limits_{t' \in T'}(x^c_{e,t'} x^{c,s}_{e,t'}+x^d_{e,t'} x^{d,s}_{e,t'}) \geq SoC^{\textit{req}}.\label{eq:EV_con3}
	\end{eqnarray}
	In each $t’+1$ the SoC is calculated as:
	\begin{eqnarray}
		SoC_{t'+1} = SoC_{t'} + \phi (x^c_{e,t'} x^{c,s}_{e,t'}+x^d_{e,t'} x^{d,s}_{e,t'}).
	\end{eqnarray}
	Also, the value of SoC should be positive and less than battery capacity. Thus, the SoC is constrained by:
	\begin{eqnarray}
		0 \leq SoC_{t'} + \phi (x^c_{e,t'} x^{c,s}_{e,t'}+x^d_{e,t'} x^{d,s}_{e,t'}) \leq B_e.
		\label{eq:EV_con4}
	\end{eqnarray}
	
	\subsection{EVCS Operation Cost Modelling}
	
	The EVCS operation mainly focuses on receiving the maximal revenue and the major profit is from EV charging schedule. The EVCS operators can obtain more profits by purchasing power from the idle charged EVs to charge other EVs for which the charging demand is not yet satisfied, if the EV discharging price is lower than the price from power grid. To maximize the benefits, the EVCS operator should perform the charging/discharging schedule to identify the EV related price at each time slot. In other words, the decision making for EV charging/discharging amounts in the business hours is the main factor to affect the EVCS revenue. 
	
	\begin{figure*}[ht!]
		\centering
		\includegraphics[scale=0.58]{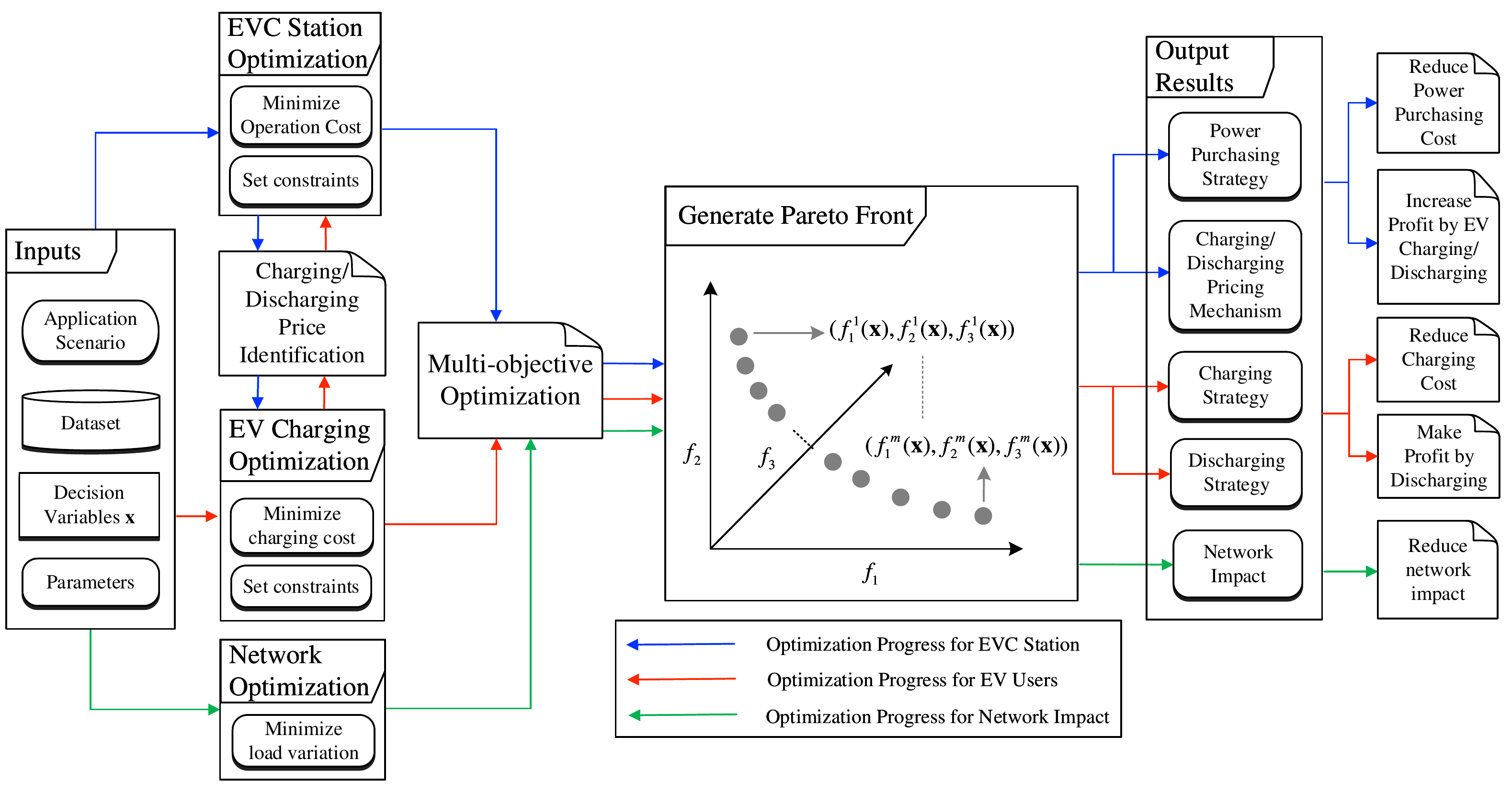}
		\caption{The proposed multi-objective electric vehicle charging/discharging schedule framework}
		\label{overall_framework}
	\end{figure*} 
	
	The profit involves the EV charging revenue and energy purchasing cost. In the V2G scenario, the EVCS purchase energy from both spot market and discharging EV fleets. To maximize the profit is to minimize the cost, modelled as:
	\begin{eqnarray}
		&\min\ f_2(X^c, X^d) = &\sum\limits_{t \in T} [(l_t+X^c_{t} + X^d_{t})_+\lambda_t^s-X^d_{t} \lambda_t \nonumber \\
		&&- X^c_{t}\lambda_t].
	\end{eqnarray}
	where $X^c_{t} \lambda_t$ is the EV charging revenue in time slot $t$. $(l_t+ X^c_{t}+X^d_{t})_+\lambda_t^s$ and $X^d_{t} \lambda_t$ are the energy purchasing cost from power grid and discharging EVs, respectively.
	
	In time slot $t$, the total charging demand and discharging energy are from all EV charging fleets and discharging fleets, respectively. The $X^c_t$ and $X^d_t$ can be calculated as:
	\begin{eqnarray}
		X^c_t = \sum\limits_{j \in C} x^c_{j,t} x^{c,s}_{j,t}, \quad x^{c,s}_{j,t}\geq 0 \quad \& \quad X^c_t \geq 0,\\
		X^d_t = \sum\limits_{i \in D} x^d_{i,t} x^{d,s}_{i,t}, \quad x^{d,s}_{i,t} \leq 0 \quad \& \quad X^d_t \leq 0.
	\end{eqnarray}
	Considering boundary constraints, the EVCS optimization can be alternatively defined as follow: 
	\begin{eqnarray}
		&&\hspace{-0.7cm}\min f_2(x^c,x^{c,s},x^d,x^{d,s}) = \sum\limits_{t \in T} [(l_t+\sum\limits_{j \in C} x^c_{j,t} x^{c,s}_{j,t} \nonumber\\
		&&\hspace{3.3cm}+\sum\limits_{i \in D} x^d_{i,t} x^{d,s}_{i,t})_+\lambda_t^s -\sum\limits_{i \in D} x^d_{i,t} x^{d,s}_{i,t} \lambda_t \nonumber\\
		&&\hspace{3.3cm} - \sum\limits_{j \in C} x^c_{j,t} x^{c,s}_{j,t}\lambda_t], \label{eq:EVC_obj}\\
		&&\hspace{-0.3cm}s.t.\ X^{\textit{min}}_t \leq (l_t + \sum\limits_{j \in C} x^c_{j,t} x^{c,s}_{j,t} + \sum\limits_{i \in D} x^d_{i,t} x^{d,s}_{i,t})_+ \leq X^{\textit{max}}_t, \label{eq:EVC_con1}\\
		&&\hspace{0.3cm}\sum\limits_{j \in C} x^c_{j,t} x^{c,s}_{j,t} \geq 0,\ \forall j \in C,\ \forall t \in T,\label{eq:EVC_con2}\\ 
		&&\hspace{0.3cm}\sum\limits_{i \in D} x^d_{i,t} x^{d,s}_{i,t} \leq 0,\ \forall i \in D,\ \forall t \in T.\label{eq:EVC_con3}
	\end{eqnarray}
	where Eq.~\ref{eq:EVC_con1} constrains the amount of energy transmission in time slot $t$ in the power grid. Eq.~\ref{eq:EVC_con2} and Eq.~\ref{eq:EVC_con3} constrain the EV charging and discharging power, respectively.
	
	A quadratic cost model~\cite{walters1993genetic} is applied to describe electricity purchasing cost in a time slot, defined as:
	\begin{eqnarray}
		S(l_t) = \alpha + \beta l_t + \gamma l_{t}^2. \label{eq:pricing}
	\end{eqnarray}
	where $\alpha,\ \beta,\ \gamma$ are cost coefficients, and $\alpha,\ \beta,\ \gamma \geq 0$. $l_t$ denotes the volume of purchased electricity. Thus, for EV charging/discharging market, the EV charging/discharging price can be formulated as follow:
	\begin{eqnarray}
		&&\lambda_t = \lambda^s_t + \alpha(l_t + X^c_t + X^d_t)^2 + \beta(l_t + X^c_t + X^d_t)\nonumber \\
		&&\hspace{0.4cm}= \lambda^s_t + \alpha(l_t + \sum\limits_{j \in C} x^c_{j,t} x^{c,s}_{j,t} + \sum\limits_{i \in D} x^d_{i,t} x^{d,s}_{i,t}) \nonumber \\
		&&\hspace{0.7cm}+ \beta(l_t + \sum\limits_{j \in C} x^c_{j,t} x^{c,s}_{j,t} + \sum\limits_{i \in D} x^d_{i,t} x^{d,s}_{i,t})^2.\label{eq:pricing_revised}
	\end{eqnarray}
	The EV charging/discharging pricing mechanism can be applied to both EV user charging scenario as $\lambda_{t'}$ and EVCS operation application as $\lambda_{t}$.
	
	\subsection{Network Impact}
	
	Minimizing the impact on the distribution network is to control the charging/discharging demand in each period. The net power exchange profile with the grid should be flattened as much as possible to allow an optimized generation dispatch and the stable grid operation~\cite{das2020multi}. The variation of the power exchange within the grid is quantified as the network impact. The impact of the grid net exchange is modelled as: 
	\begin{eqnarray}
		\min\ && f_3(x^c, x^{c,s}, x^d, x^{d,s}) \nonumber\\ 
		&& = \sum\limits_{t \in T}[(l_t+\sum\limits_{j \in C} x^c_{j,t} x^{c,s}_{j,t} + \sum\limits_{i \in D} x^d_{i,t} x^{d,s}_{j,t})^2].\label{obj:3}  
	\end{eqnarray}
	
	Eq.~\ref{obj:3} is calculated to evaluate the network impact, where the constraint energy transmission is same as Eq.~\ref{eq:EVC_con1}.  
	
	\section{Proposed Method}\label{proposed_method}
	
	This section firstly introduces the proposed MOEVCS framework and the core elements. Then, the developed constraint MVMOEA used to implement the proposed MOEVCS is detailed.   
	
	\subsection{Framework}
	
	Fig.~\ref{overall_framework} depicts the proposed framework for addressing the EV charge/discharge schedule problem. The main components include the inputs, multi-objective functions, multi-objective optimization strategy, Pareto front generated, the output results and the extended analysis from outputs. 
	
	The inputs consist of the application scenario, the dataset, the decision variables $\textbf{x}$, the parameters in the scenario and methods. As illustrated in Section.~\ref{problem_modeling}, we focus on investigating the coordinated charging and discharging schedule for different EVs over $T$ periods, where the EVs have different arriving times, departure times, charging requests, and SoC. The objectives and constraints include EVCS optimization, EV charging optimization, and network impact formulated. EVCS and EV users mainly focus on minimizing their cost using Eq.~\ref{eq:EVC_obj} and Eq.~\ref{eq:EV_obj} to maximize their benefits. The network impact optimization (Eq.~\ref{obj:3}) mainly aims at reducing the energy spikes by minimizing the load variation over the investigated time period to keep the distribution network stable. EVCS and EV users optimization influences the charging/discharging price since TOU tariff is decided by the total load at different time stamps according to Eq.~\ref{eq:pricing_revised}. The objectives $f_1$, $f_2$, and $f_3$ and the related constraints including Eq.~\ref{eq:EV_con3} and Eq.~\ref{eq:EV_con4} are fed into a constraint MOEA to obtain the trade-off solutions for $\textbf{x}$ and generate the evolved Pareto Front (PF). The generated PF includes a set of good solutions that balance the competing objectives. Finally, the results related to the power purchase for the EVCS, the TOU tariff, the charging and discharging schedule for the EV users, and the network impact can be obtained from the PF and the solutions. Further analysis can also be performed, such as increasing the benefits of different stakeholders and reducing the network impact.  
	
	\subsection{Implementation}
	To implement the proposed MOEVCS, we develop a constraint MVMOEA for addressing the problem formulated. We firstly describe the constraint MOO problems (CMOOPs). Then the elements in decision variables $\textbf{x}$ are explained and the proposed MVMOEA is presented. 
	
	\subsubsection{Constraint Multi-objective Optimization}
	In this scenario, any EV $e$ should reach $SoC^\textit{req}$ (i.e., $100\%$) before it leaves the EVCS, which means that it should be fully charged with $B$. Eq.~\ref{eq:EV_con3} can be replaced by: 
	\begin{eqnarray}
		SoC^{\textit{arr}} + \phi \sum\limits_{t' \in T'}(x^c_{e,t'} x^{c,s}_{e,t'}+x^d_{e,t'} x^{d,s}_{e,t'}) - SoC^{\textit{req}} = 0.\label{eq:EV_con5}
	\end{eqnarray}
	We use $h_e(\textbf{x}), e = 1,..., N_\textit{user}$ to represent Eq.~\ref{eq:EV_con5} and $h_e(x) = 0$. Similarly, Eq.~\ref{eq:EV_con4} can be replaced by:
	\begin{eqnarray}
		&&-SoC_{t'} - \phi (x^c_{e,t'} x^{c,s}_{e,t'}+x^d_{e,t'} x^{d,s}_{e,t'}) \leq 0,  \\
		&&SoC_{t'} + \phi (x^c_{e,t'} x^{c,s}_{e,t'}+x^d_{e,t'} x^{d,s}_{e,t'}) - B_e\leq 0.  \label{eq:EV_con6}
	\end{eqnarray}
	We use $g_{e, t'}(\textbf{x})$ to represent the unequal constraints so that $g_{e, t'}(\textbf{x}) \leq 0$. According to CMOOPs, we can define our problem as follows:
	\begin{align}
		\label{MOP}
		\left\{\begin{array}{rl}
			\min & \textbf{F}(\textbf{x})=(f_1(\textbf{x}), f_2(x),..., f_m(\textbf{x}))^T,  \\ 
			s.t. & g_j(\textbf{x}) \leq 0, \ j = 1,..., p, \\
			& h_j(\textbf{x}) = 0, \ j = p+1,..., p+q, \\
			& x\in \Omega,\ m = 3.   
		\end{array}\right.
	\end{align}
	where $\Omega \subset \mathbb{R}^d$ represents the decision space and $\textbf{x}=(x_1, x_2,..., x_d) \in \Omega$ is a decision variable with respect to a solution for a specific CMOOP. $\textbf{F}(\textbf{x}): \Omega \to \mathbb{R}^m$ denotes the $m$-dimensional objective vector of the solution $\textbf{x}$. To simplify this problem, the constraints $h_j(\textbf{x}) = 0$ can be represented by $|h_j(\textbf{x})|-\epsilon \leq 0$, where $\epsilon$ is a very small tolerance value to relax equality constraints. The constraint violation of $\textbf{x}$ on the $j^{th}$ constraint can be obtained via: 
	\begin{align}
		\label{CV}
		CV_j(\textbf{x}) =  \left\{\begin{array}{rl}
			&\hspace{-0.4cm} max(0, g_j(\textbf{x})), \ j = 1,..., p, \\
			&\hspace{-0.4cm} max(0, |h_j(\textbf{x})|-\epsilon), \ j = p+1,..., p+q.
		\end{array}\right.
	\end{align}
	Therefore, $\textbf{x}$ is a feasible solution if $CV(\textbf{x}) = 0, CV(\textbf{x}) = \sum_{j=1}^{p+q}CV_j(\textbf{x})$. For any two solutions $\textbf{x}_A$ and $\textbf{x}_B$,  $\textbf{x}_A$ is better than $\textbf{x}_B$ if:   
	\begin{align}
		\label{feasible_solution}
		& \hspace{-5.5cm} CV(\textbf{x}_A) < CV(\textbf{x}_B),\ or\\
		\left\{\begin{array}{rl}
			& f_i(\text{\textbf{x}}_A) \leq f_i(\text{\textbf{x}}_B),\ \forall i\in \{1,..., m\}, \\
			& f_j(\text{\textbf{x}}_A) < f_j(\text{\textbf{x}}_B),\ j\in \{1,..., m\}, \\
			& CV(\textbf{x}_A) = CV(\textbf{x}_B).
		\end{array}\right.
	\end{align}
	
	A solution $\textbf{x}^* \in \Omega$ is called Pareto optimum if there is no other solution that dominates $\textbf{x}^*$. The set of all Pareto optimal solutions is called Pareto optimal set (PS) and Pareto optimal front is defined as the objective vectors of the solutions in the PF. Therefore, to solve a CMOOP is to find its PS. 
	\begin{figure}[ht!]
		\centering
		\includegraphics[scale=0.65]{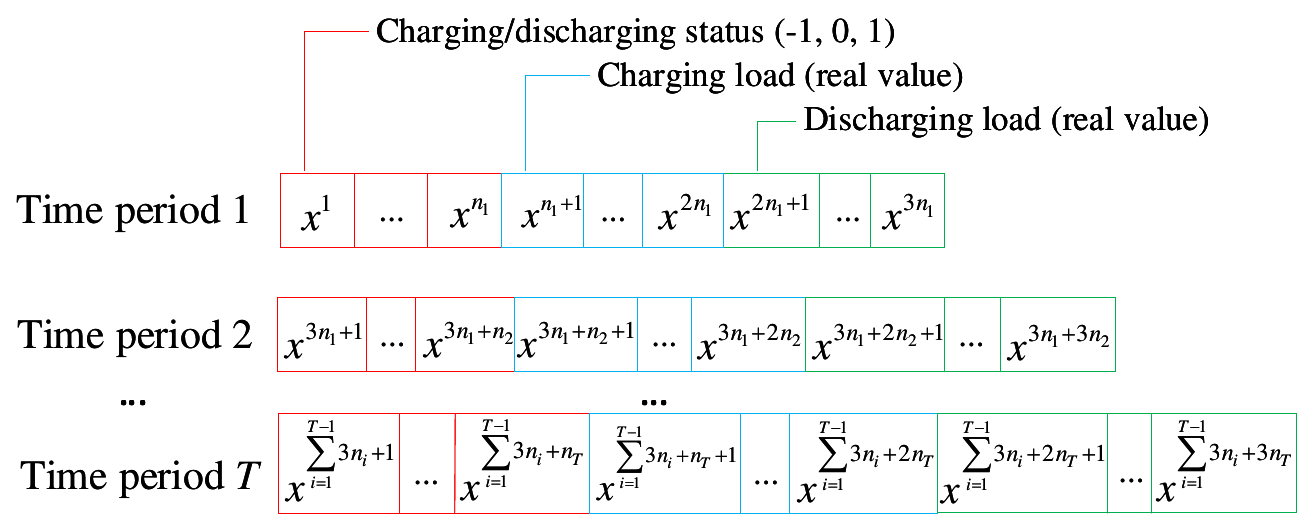}
		\caption{Representation of a solution $\textbf{x} = (x^1, x^2, ..., x^d), d = \sum_{i=1}^{T} 3n_i$ from time period 1 to $T$, where $d$ denotes the number of decision variables, $n_i$ represents the number of EV users having the charging/discharging requests during $i$th time period.}
		\label{dimension_structure}
	\end{figure} 
	
	\subsubsection{Encoding and Decoding of The Solutions}
	\label{encoding_decoding} 
	
	According to the problem modelling in Section.~\ref{problem_modeling}, the decision variables $\textbf{x}$ in the formulated objective functions include the state of charging $x^{c,s}$ and discharging $x^{d,s}$, the charging load $x^c$ and discharging load $x^d$. As explained in Section.~\ref{EV_user_modeling}, the state of charging and discharging cannot be simultaneously active at the same period for one EV, so $x^{c,s}$ and $x^{d,s}$ can be represented by an integer decision variable $x^{s} \in \{-1, 0, 1\}$ during the evolution process. $x^{s} = -1$, $x^{s} = 0$, and $x^{s} = 1$ lead to the discharging, idle, and charging state active, i.e., $\{x^{d,s} = -1, x^{c,s} = 0\}$, $\{x^{d,s} = 0, x^{c,s} = 0\}$, and $\{x^{d,s} = 0, x^{c,s} = 1\}$. $x^c$ and $x^d$ are continuous values. 
	
	\begin{algorithm}[h!]
		\footnotesize
		\caption{The developed MVMOEA}	\label{algorithm_MOEVCS}
		\SetAlgoLined
		\KwIn{A MOP with its solution space $\Omega$, objective function $(f_1, f_2, f_3)$, constraints in Eq.~\ref{MOP}, constraint violation in Eq.~\ref{CV}, population size $\textit{ps}$, crossover rate $\textit{CR}$, mutation probability $\textit{MP}$, $\textit{MaxGen}$, $d$, $\textit{Gen} = 0$, $x^s_b$, $x^{c}_b$, $x^{d}_b$: bounds of $x^s$, $x^{c}$, $x^{d}$ in $\textbf{x}$}
		\KwOut{Pareto front solutions}
		\textbf{Step 1) Initialization:}\\
		\textbf{Step 1.1)} Randomly generate an initial population $\{\textbf{x}_1,..., \textbf{x}_\textit{ps}\}$ with uniform distribution in search space $\Omega$, then for each individual $\textbf{x}_i, i=1,..., ps$, decodes each dimension into its real space according to Section.~\ref{encoding_decoding} \\
		\textbf{Step 1.2)} Evaluate each candidate solution $\textbf{x}_i, i=1,..., \textit{ps}$ to obtain its objective fitness $f_j^i, j = 1,..., 3, i = 1,..., \textit{ps}$ and constraint violation $CV(\textbf{x}_i)$ according to Eq.~\ref{feasible_solution}, $\textit{Gen} = \textit{Gen} + 1$\\   
		\textbf{Step 1.3)} \textbf{Fast-non-dominated-sorting}~\cite{deb2002fast}\\
		\textbf{Step 2) Evolution:}\\
		\While{$\textit{Gen} < \textit{MaxGen}$}{
			\textbf{Step 2.1) Selection:} tournament selection according to Eq.~\ref{feasible_solution} to generate $\textit{ps}$ children\\
			\textbf{Step 2.2) Crossover:} crossover operation with $\textit{CR}$ using Simulated Binary Crossover \cite{deb2007self} on the newly generated individuals\\
			\textbf{Step 2.3) Mutation:} Polynomial Mutation with $\textit{MP}$\\
			\textbf{Step 2.4) Evaluation:} decoding according to Section.~\ref{encoding_decoding} and evaluate each newly generated individuals according to \textbf{Step 1.2}, $\textit{Gen} = \textit{Gen} + 1$\\
			\textbf{Step 2.5) Fast-non-dominated-sorting:} combine the parents and children, perform non-dominated-sorting according to ~\cite{deb2002fast}, and select $\textit{ps}$ individuals as the parents\\
		}						
	\end{algorithm}

	For any potential solution $\textbf{x} = (x_1, x_2,..., x^d)$ with $d = \sum_{i=1}^{T}3n_i$ decision variables, the representation of each dimension is illustrated in Fig.~\ref{dimension_structure}. For any time period $i\in [1, T]$ with $n_i$ EV users that may have the charging/discharging requests, there are $3n_i$ decision variables to be optimized, consisted of charging, discharging or idle states, charging load, and discharging load. Therefore, the decision variables consist of discrete integers and continuous values.
	
	\begin{table*}[ht]
		\centering
		\caption{The charging/discharging request for each problem set during the investigated time stamps}
		\footnotesize
		\label{set_distribution}
		\setlength{\tabcolsep}{3pt}
		\begin{tabular}{|m{0.5cm}|m{0.35cm}|m{0.35cm}|m{0.35cm}|m{0.35cm}|m{0.35cm}|m{0.35cm}|m{0.35cm}|m{0.35cm}|m{0.35cm}|m{0.35cm}|m{0.35cm}|m{0.35cm}|m{0.35cm}|m{0.35cm}|m{0.35cm}|m{0.35cm}|m{0.35cm}|m{0.35cm}|m{0.35cm}|m{0.35cm}|m{0.35cm}|m{0.35cm}|m{0.35cm}|m{0.35cm}|m{0.35cm}|m{0.35cm}|m{0.35cm}|m{0.35cm}|m{0.35cm}|}
			\hline
			&01&02& 03 & 04 & 05 & 06 & 07 & 08 & 09 & 10 & 11 & 12 & 13 & 14 & 15 & 16 & 17 & 18 & 19 & 20 & 21 & 22 & 23 & 00 & 01 & 02 & 03 & 04 & 05  \\ \hline
			Set1   & \gradient{10} & \gradient{10} & \gradient{10} & \gradient{10} & \gradient{10} & \gradient{10} & \gradient{10} & \gradient{13} & \gradient{10} & \gradient{10} & \gradient{10} & \gradient{10} & \gradient{10} & \gradient{10} & \gradient{10} & \gradient{10} & \gradient{10} & \gradient{10} & \gradient{10} & \gradient{10} & \gradient{10} & \gradient{20} & \gradient{20} & \gradient{17} & \gradient{10} & \gradient{10} & \gradient{10} & \gradient{10} & \gradient{10}  \\ \hline
			Set2   & \gradient{15} & \gradient{15} & \gradient{20} & \gradient{20} & \gradient{20} & \gradient{20} & \gradient{20} & \gradient{23} & \gradient{15} & \gradient{15} & \gradient{10} & \gradient{10} & \gradient{10} & \gradient{10} & \gradient{10} & \gradient{10} & \gradient{10} & \gradient{10} & \gradient{10} & \gradient{10} & \gradient{10} & \gradient{20} & \gradient{20} & \gradient{17} & \gradient{10} & \gradient{10} & \gradient{10} & \gradient{10} & \gradient{10}  \\ \hline
			Set3   & \gradient{10} & \gradient{10} & \gradient{10} & \gradient{10} & \gradient{10} & \gradient{10} & \gradient{10} & \gradient{13} & \gradient{10} & \gradient{15} & \gradient{15} & \gradient{15} & \gradient{15} & \gradient{15} & \gradient{15} & \gradient{15} & \gradient{15} & \gradient{15} & \gradient{15} & \gradient{15} & \gradient{15} & \gradient{25} & \gradient{25} & \gradient{22} & \gradient{15} & \gradient{10} & \gradient{10} & \gradient{10} & \gradient{10}  \\ \hline
			Set4   & \gradient{10} & \gradient{10} & \gradient{10} & \gradient{10} & \gradient{10} & \gradient{10} & \gradient{10} & \gradient{20} & \gradient{10} & \gradient{10} & \gradient{10} & \gradient{10} & \gradient{10} & \gradient{10} & \gradient{10} & \gradient{10} & \gradient{15} & \gradient{15} & \gradient{15} & \gradient{15} & \gradient{15} & \gradient{30} & \gradient{30} & \gradient{20} & \gradient{15} & \gradient{15} & \gradient{15} & \gradient{15} & \gradient{15}  \\ \hline
		\end{tabular}
	\end{table*}

	With $n_i$ EVs that probably perform charging or discharging requests at time period $i\in [1, T]$, the state for each EV user is $x_{i, j}^{s}, j \in[1, n_i]$ ($x^{\sum_{k = 1}^{i-1}3n_k + j}$ in Fig.~\ref{dimension_structure} if $i >1$, else $x^{j}$), and $x_{i, j}^{s} \in\{-1, 0, 1\}$. $x_{i, j}^{s} = -1$ means that the $i$th EV has the discharging activity, while $x_{i, j}^{s} = 1$ represents the charging activity. $x_{i, j}^{s} = 0$ means that there is no request. To simplify this, we set $x_{i, j}^{d,s} = 0, x_{i, j}^{c,s} = 1$ when $x_{i, j}^{s} = 1$ and $x_{i, j}^{d,s} = -1, x_{i, j}^{c,s} = 0$ when $x_{i, j}^{s} = -1$. $x_{i, j}^{d,s} = 0, x_{i, j}^{c,s} = 0$ when $x_{i, j}^{s} = 0$. In this case, $(x^{\sum_{k = 1}^{i-1}3n_k + 1},..., x^{\sum_{k = 1}^{i-1}3n_k + n_{i}})$, $(x^{\sum_{k = 1}^{i-1}3n_k + n_{i} + 1},..., x^{\sum_{k = 1}^{i-1}3n_k + 2n_{i}})$, and $(x^{\sum_{k = 1}^{i-1}3n_k + 2n_{i} + 1},..., x^{\sum_{k = 1}^{i-1}3n_k + 3n_{i}})$ (or $(x^{1},..., x^{n_{1}})$, $(x^{n_{1} + 1},..., x^{2n_{1}})$, and $(x^{2n_{1} + 1},..., x^{3n_{1}})$ if $i = 1$) denote the states, the discharging and charging load for each EV user at $i$th time period, respectively. If $j$th EV at 04:00 am ($i = 4$) has the state ‘1’ ($x_{4, j}^s = 1$) and the load is 8.5kW, then $x_{4, j}^{c,s} = 1$, $x_{4, j}^{d,s} = 0$ and $x_{4, j}^{c}$ = 8.5kW. $x_{4, j}^{d}$ is inactive due to $x_{4, j}^{d,s} = 0$.
	
	\subsubsection{Constraint Non-dominated Sorting Genetic Algorithm II}
	
	Over the past decade, a number of MOEAs~\cite{zhou2011multiobjective} have been introduced for addressing MOOPs and finding solutions that makes a trade-off between the competing objectives. Since EAs work with a population of solutions, a simple EA can be extended to maintain a diverse set of solutions. Since MOEAs are mainly proposed for solving continuous optimization problems, we develop a mixed-variable MOEA (MVMOEA) to implement our proposed MOEVCS. We employ one popular MOEV, i.e., NSGA-II~\cite{deb2002fast}, as the MOO solver to address our problem. By integrating NSGA-II\footnote{https://github.com/msu-coinlab/pymoo} with our problem, the pseudocode of MVMOEA is presented in Algorithm.~\ref{algorithm_MOEVCS}.  
	
	Given the inputs and the problem dimension $d$, MVMOEA first initializes $\textit{ps}$ individuals fixed within continuous ranges. Each solution is decoded into the dimension representation as illustrated in Fig.~\ref{dimension_structure}. Then, all the decoded solutions are evaluated on the objective functions $f_1, f_2, f_3$ and the constraints in Eq.~\ref{MOP}. The constraint violations are obtained via Eq.~\ref{CV}. According to the feasible solution condition in Eq.~\ref{feasible_solution}, fast non-dominated sorting (NDS)~\cite{deb2002fast} is applied to obtain the PF. During the evolution process, the operations such as selection, crossover and mutation are same with the general GA. Before evaluating any newly generated child, it is decoded according to Fig.~\ref{dimension_structure} for easily applying to Eq.~\ref{eq:EV_obj}, Eq.~\ref{eq:EVC_obj} and Eq.~\ref{obj:3}. After evaluating the generated children, the children and parents are combined and NDS is performed according to the fitness values over all objective functions and the constraint violations. Finally, $ps$ individuals are selected as the new parents for the evolution of the next generation.

	\section{Experiments}\label{experiments} 
	To demonstrate the effectiveness of the proposed MOEVCS, the following hypotheses are verified, explored, and analysed. 
	
	\begin{itemize}
		\item We design five different baselines to assess the performance including i) average charging request on the charging period, ii) first-come-first-served strategy, and iii, iv,  v) single objective strategy by minimizing the network impact (i.e., Eq.~\ref{obj:3}), EVCS cost (i.e., Eq.~\ref{eq:EVC_obj}), and user cost (i.e., Eq.~\ref{eq:EV_obj}), separately. 
		
		\item The analysis of the users' cost of and the EVCS's benefit (i.e., cost) further verifies MOEVCS. The total load schedule after coordinated charging/discharging can balance the network stability.  
		
		\item Different application scenarios explored show the effectiveness of the proposed MOEVCS on the coordinated charging/discharging problem. 
	\end{itemize}

	\begin{table}[!h]
		\centering
		\caption{Problem sets and the related information}\label{problem_info}
		\begin{tabular}{|c|cclc|}
			\hline
			Problems & \multicolumn{1}{c|}{$N_\textit{user}$} & \multicolumn{1}{c|}{$N_\textit{const}$} & \multicolumn{1}{l|}{$d$} & Request Centralization \\ \hline
			Set 1 & 40 & 600 & 960 & \begin{tabular}[c]{@{}c@{}}Stable except\\ 22:00 pm - 00:59 am\end{tabular} \\ \hline
			Set 2 & 50 & 750 & 1200 & 01:00 am - 10:59 am \\ \hline
			Set 3 & 50 & 750 & 1200 & 10:00 am - 01: 59 am   \\ \hline
			Set 4 & 50 & 750 & 1200 & 17:00 pm - 05:59 am \\ \hline
		\end{tabular}
	\end{table}

	\subsection{Data Description}
	We apply the EV trajectory and office building dataset as in~\cite{li2017data}. The EV trajectory dataset is collected from 490 EVs from Shenzhen city in November 2013. The EV charging time period, demand, and the start SoC is statistically summarized from the original dataset. The charging period for each EV fleet is set to 8 hours. Based on the EVs investigated, the overall charging period is from 01:00 am to 05:59 am in the next day (i.e., 29 hours). The office building dataset contains the hourly base load in the same city and time period. The parking lot is applied as the EVCS and the building base load is considered as the operation load.

	We design four different scenarios according to the charging and discharging activities of the customers based on the same base load of the EVCS. Table.~\ref{set_distribution} illustrates the centralization of charging/discharging requests over different problem sets. There are some significant peak hours for each set due to the real application. Table.~\ref{problem_info} presents the information related to the number of users $N_\textit{user}$, number of constraints $N_\textit{const}$, dimension $D$ and request centralization for each problem set. 
	
	\begin{table}[!h]
		\centering
		\caption{The summary of the parameters}\label{parameters}
		\begin{tabular}{|ccccc|}
			\hline
			$x_b^s$   & \begin{tabular}[c]{@{}c@{}}Capacity\\ ($B/\text{kW}$)\end{tabular} & \begin{tabular}[c]{@{}c@{}}Cost\\ ($C^B/\text{CNY}$)\end{tabular} & $\textit{ps}$ & \begin{tabular}[c]{@{}c@{}}Maximal \\ charging/discharging\\ power (kWh)\end{tabular} \\ \hline
			{[}-1,1{]} & 50 & 120000 & 1000 & 10 \\ \hline
			$k$ & $\textit{MaxGen}$ & $\textit{CR}$  & $\textit{MP}$ & Types  \\ \hline
			-1/64   & 2E+04   & 0.95   & 0.01 & Tesla Model 3  \\ \hline
			$\epsilon$ & $\beta, \gamma$ & $X^\textit{max}$ & $X^\textit{min}$ & $x_b^c/x_b^d$ \\ \hline
			0.1 & 5E-05   & 1000 & 0 & {[}0,10{]}  \\ \hline
		\end{tabular}
	\end{table}	
	\subsection{Experimental Setup}
	The experimental setups include the parameters related to EVs, the tariff structure in Eq.~\ref{eq:pricing}, SOEA implemented by SOGA, MVMOEA, and constraints. The core parameters in SOEA include the population size $\textit{ps}$, the maximal generation $\textit{MaxGen}$, crossover rate $\textit{CR}$ and mutation probability $\textit{MP}$, which are set the same as those of MVMOEA. All the parameters are summarized in Table.~\ref{parameters}. We set $\lambda_{t'}^s = \lambda_t^s = 0.2084$ over the studied time periods according to the commercial electricity price in Shenzhen\footnote{https://http://fgw.sz.gov.cn/attachment/0/804/804296/7763984.pdf}.

	\begin{figure*}[h!]\centering
		\subfigure[Set 1]
		{\centering\scalebox{0.30}
			{\includegraphics{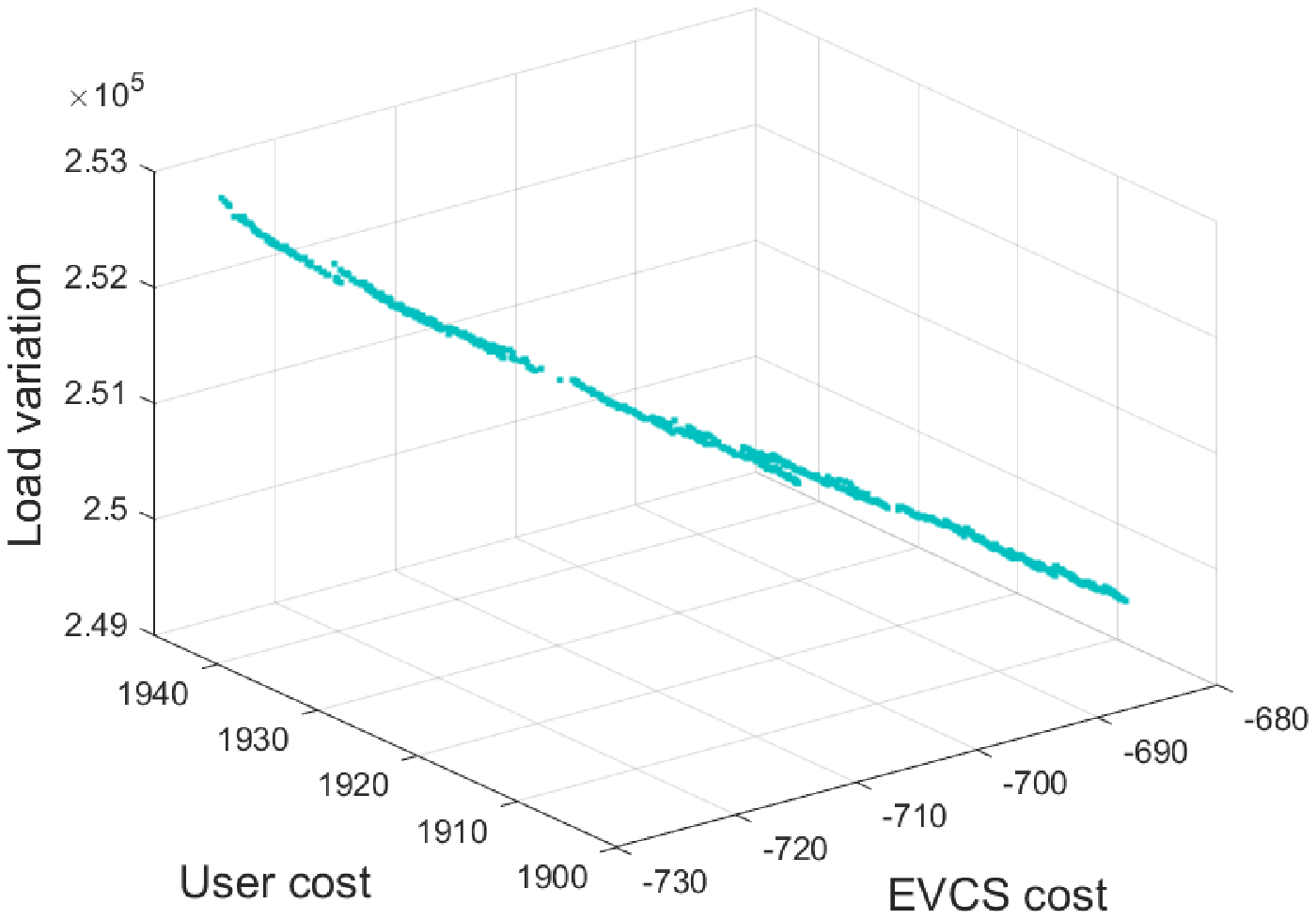}}\label{pf_set1}}\hspace{-0.3cm}
		\subfigure[Set 2]
		{\centering\scalebox{0.30}
			{\includegraphics{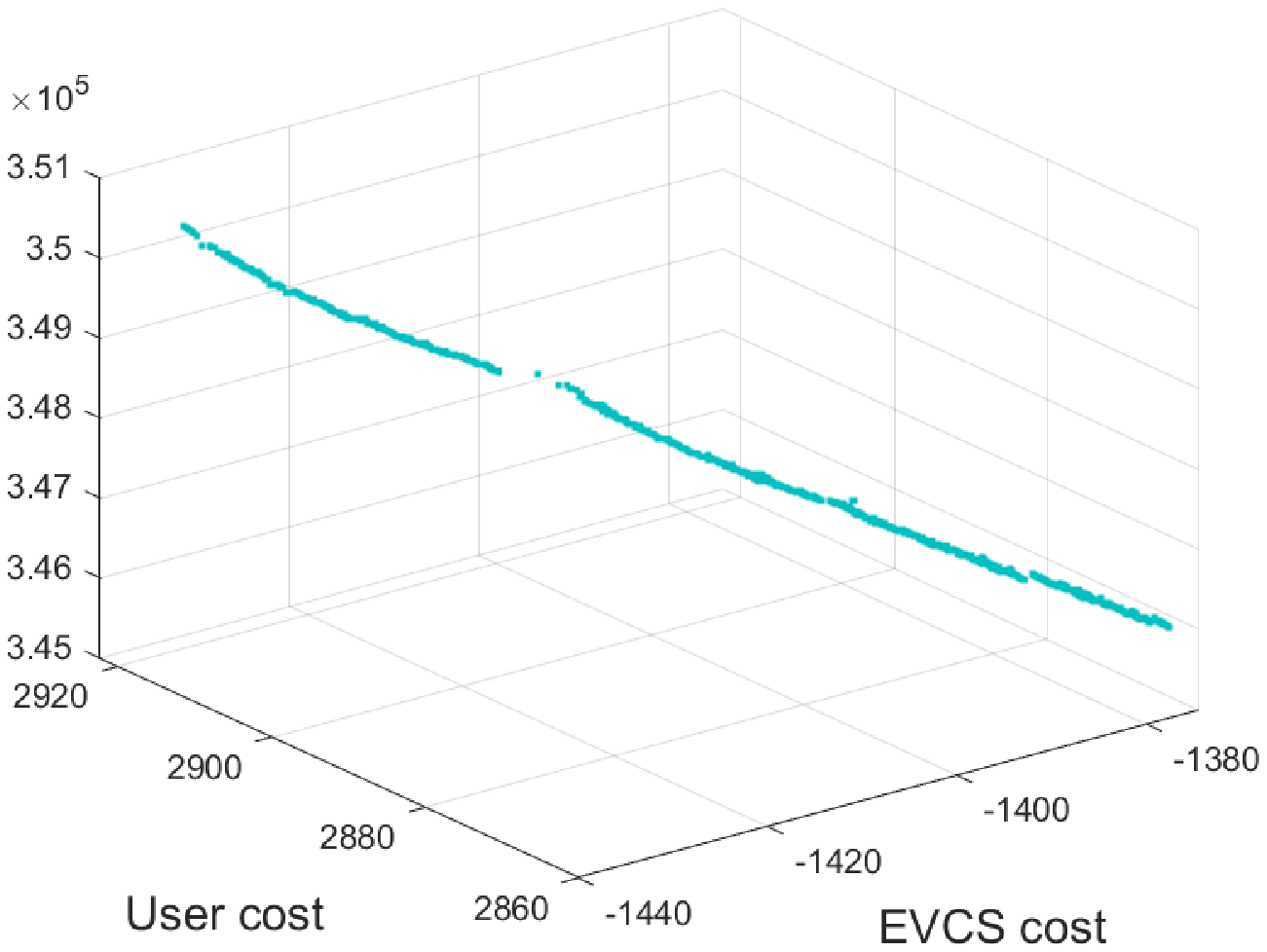}}\label{pf_set2}}\hspace{-0.3cm}
		\subfigure[Set 3]
		{\centering\scalebox{0.30}
			{\includegraphics{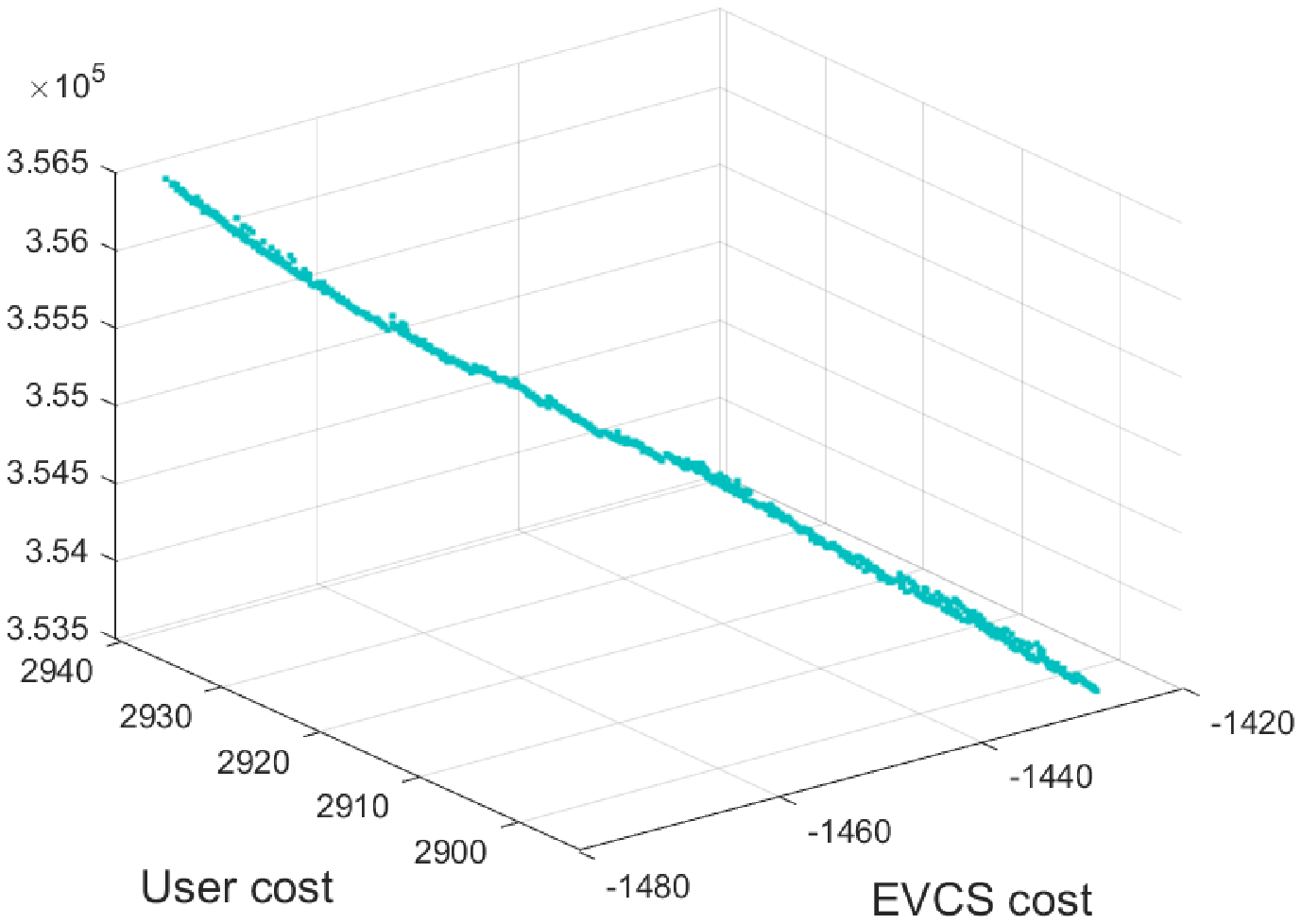}}\label{pf_set3}}\hspace{-0.2cm}
		\subfigure[Set 4]
		{\centering\scalebox{0.30}
			{\includegraphics{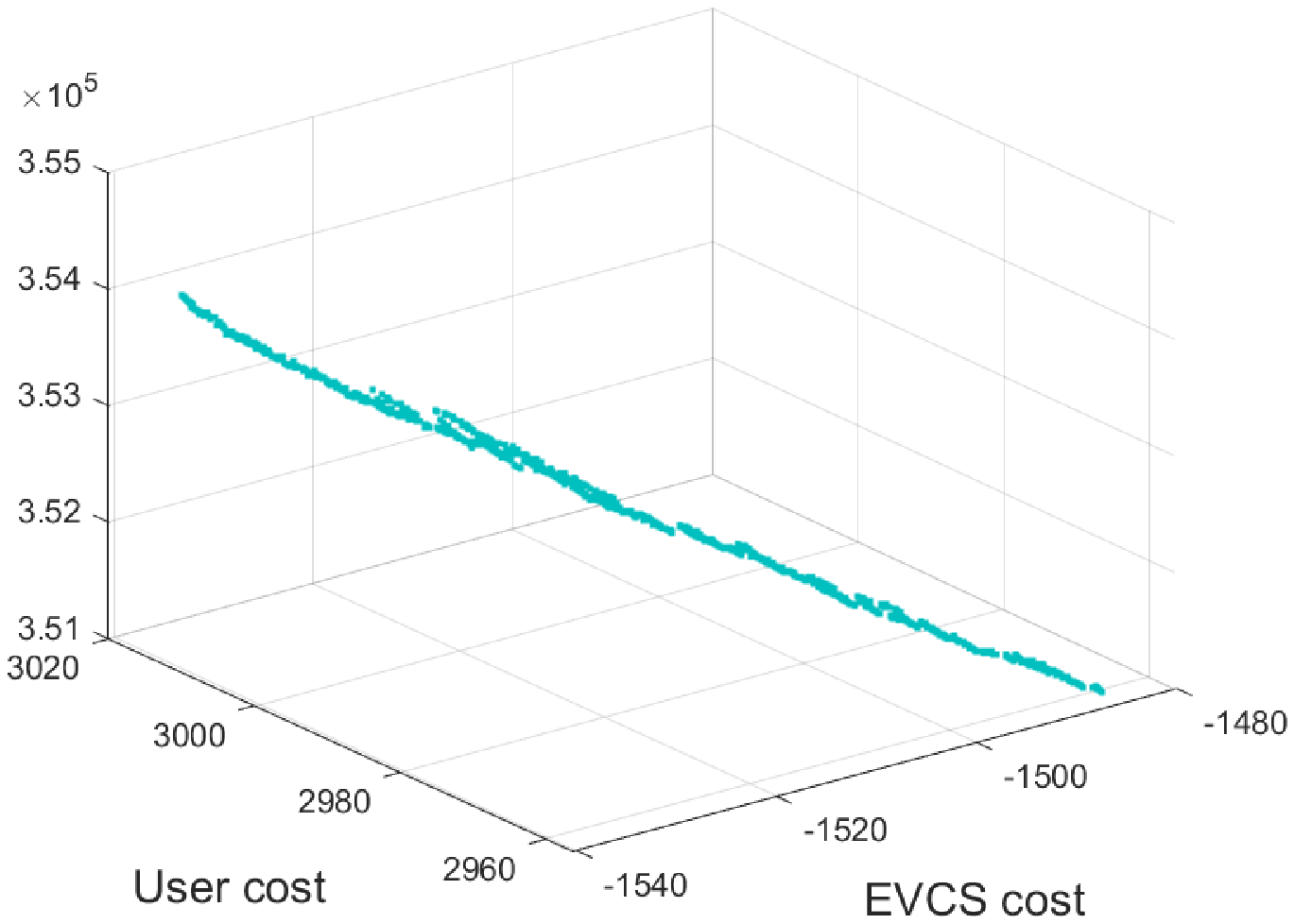}}\label{pf_set4}}
		\caption{Pareto fronts of EVCS cost, user cost and network impact over each problem set} \label{com_pf}
	\end{figure*}
	
	To demonstrate the effectiveness of MOEVCS, we design five different baselines to compare with, where there is no optimization and discharging for baselines 1 and 2 and the last three baselines are based on the SOEA by considering the charging/discharging schedule using GA.   
	\begin{itemize}
		\item Baseline 1: average charging request on the charging period, denoted as B*1;  
		\item Baseline 2: first-come-first-served strategy, denoted as B*2;
		\item Baselines 3/4/5: optimize the network impact ($f_3$), EVCS cost ($f_2$), and user cost ($f_1$) with SOEA separately, denoted as B*3/B*4/B*5.
	\end{itemize}
	$*$ represents the result of each item. For example, Bprice1 means the TOU tariff over the scheduling strategy of baseline 1. B1 without $*$ denotes the results over the relatted objective function. The results from MOEVCS include MO*MinObj2 and MO*MinObj1/3. 
	
	\begin{itemize}
		\item MO*MinObj2: one of the results on the obtained PF from MOEVCS where $f_2$ (i.e., Eq.~\ref{eq:EVC_obj}) is minimal.
		
		\item MO*MinObj1/3: one of the results on the obtained PF from MOEVCS where $f_1$ or $f_3$ (i.e., Eq.~\ref{eq:EV_obj} or Eq.~\ref{obj:3}) is minimal, given that they have the same change direction in this application scenario.  
	\end{itemize}
	
	\subsection{Results}
	This section reports and analyses the result from the baselines, SOEA, and MOEVCS from the users' cost, the EVCS's benefit, and the network impact. We first present the results of MOEVCS from the obtained PFs along with the ranges of the objective functions to demonstrate how MOEVCS balance the trade-off relationship between different competing objectives. Then, the results from MOEVCS are compared with the baselines. Finally, we analyse the total load after the coordinated charging/discharging schedule and TOU tariff. 
	
	\subsubsection{Results of MOEVCS}
	Fig.~\ref{com_pf} illustrates the trade-off relationship between the benefits of the EVCS and users as well as the network impact, quantified by minimizing the EVCS cost, users' cost, and load variation. The user cost and load variation have the same change direction over all investigated problem sets. When the users have less cost with the charging and discharging schedule, the network tends to be more stable, but the cost of the EVCS increases indicating less benefit. The decrease cost of the EVCS (more benefit) leads to the increased user cost and the larger network impact. 
	\begin{figure}[!ht]\centering
		\subfigure[EVCS cost]
		{\centering\scalebox{0.21}
			{\includegraphics{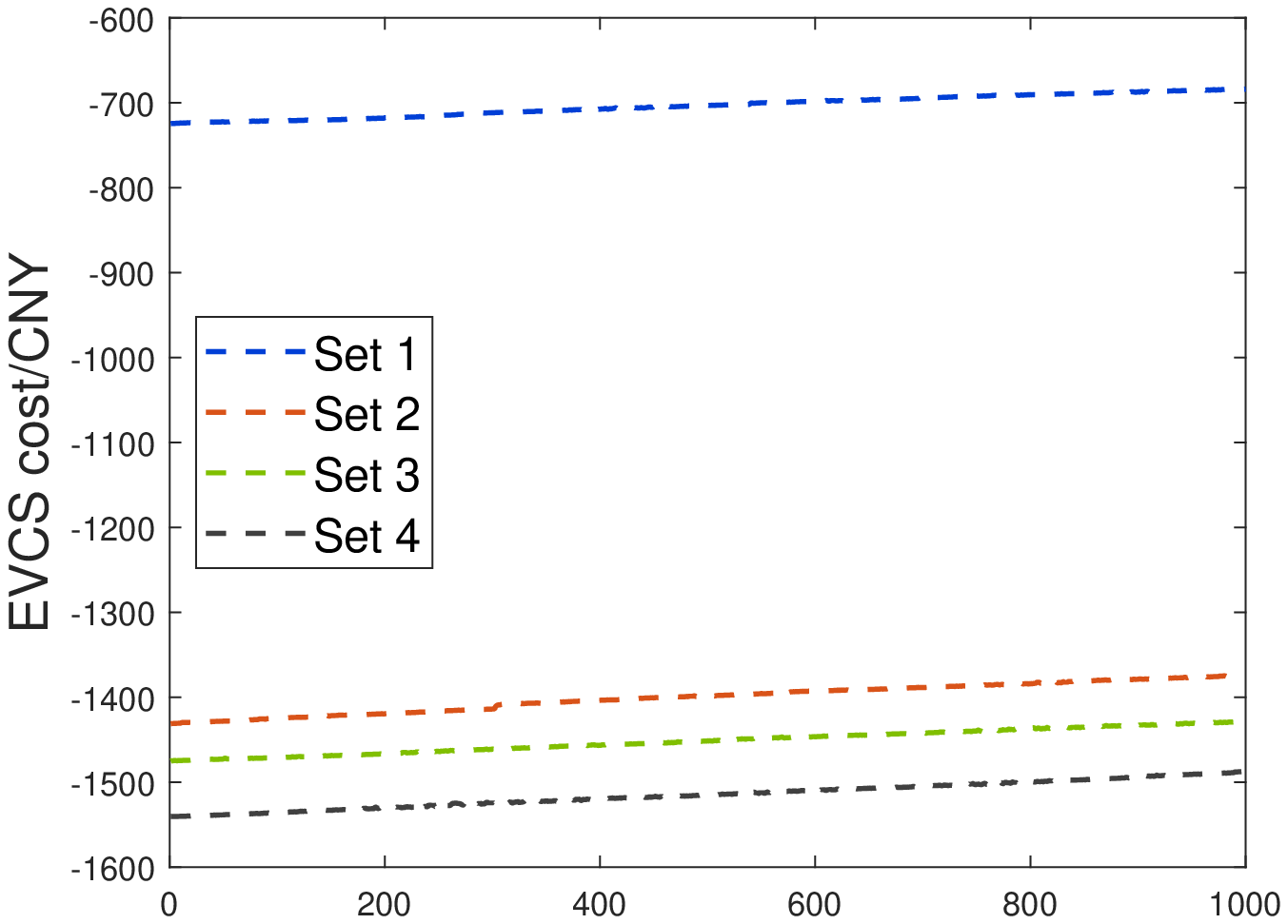}}\label{EVCScost}}
		\subfigure[User cost]
		{\centering\scalebox{0.21}
			{\includegraphics{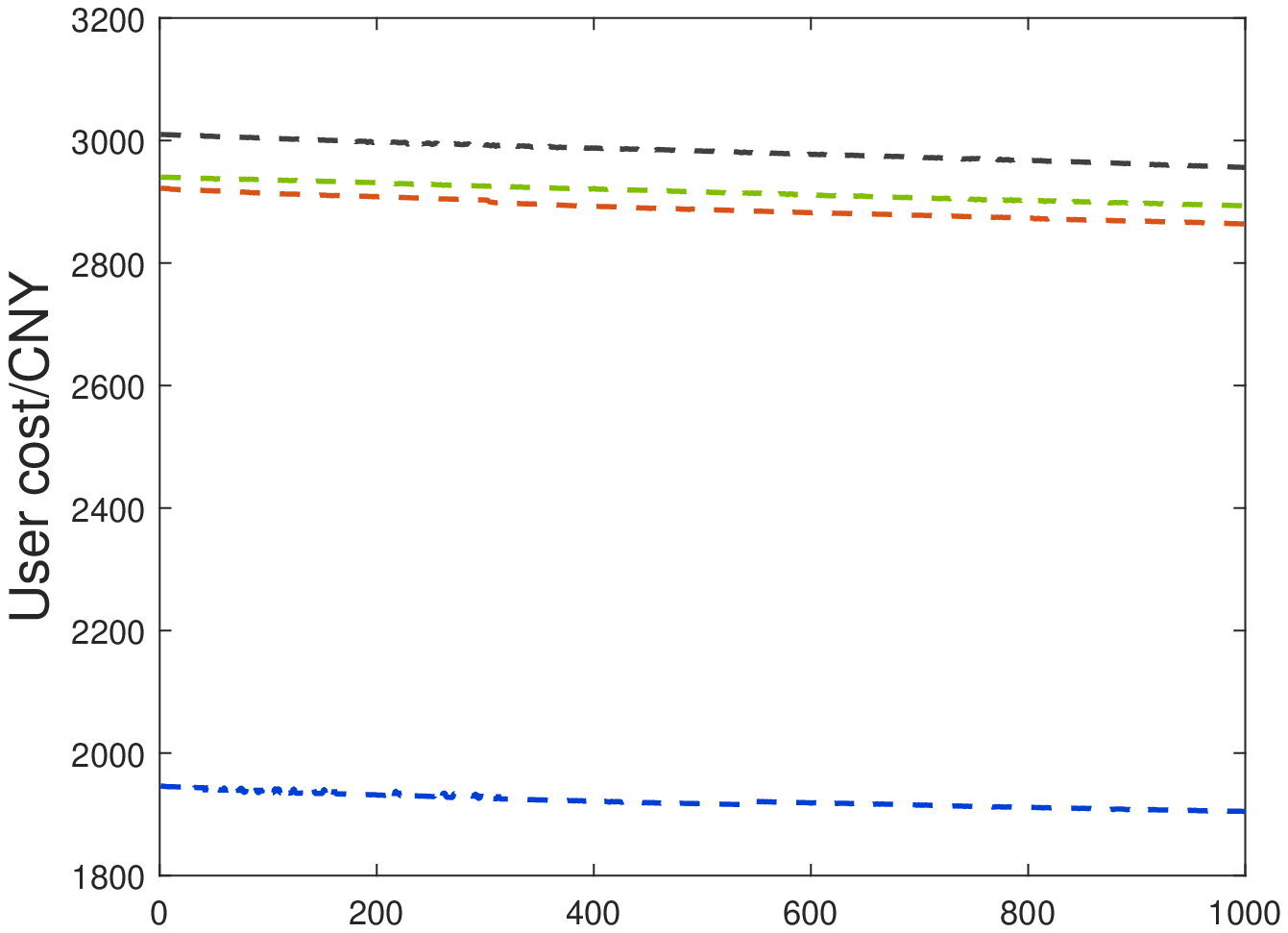}}\label{Usercost}}
		\subfigure[Load variation]
		{\centering\scalebox{0.21}
			{\includegraphics{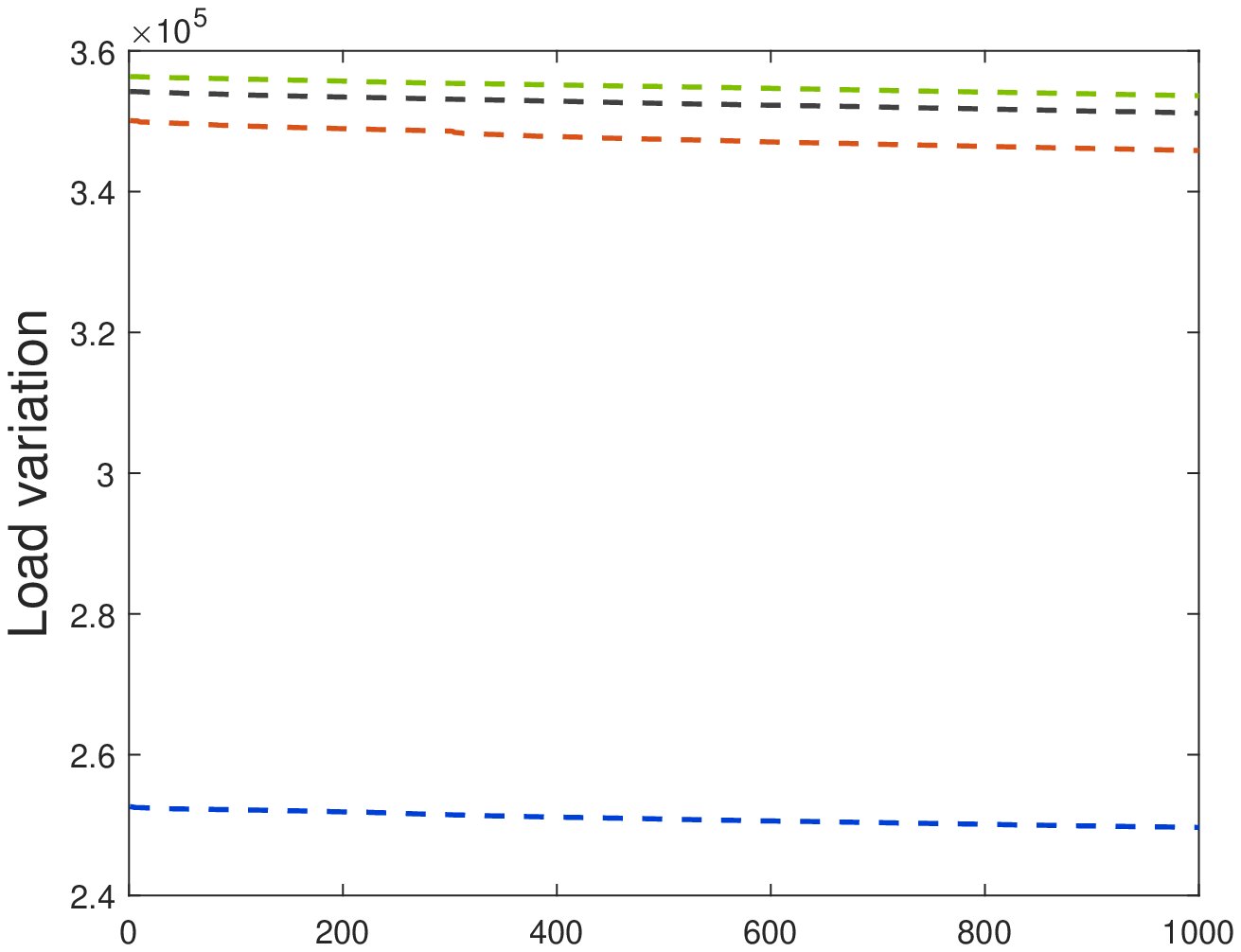}}\label{Loadvariation}}\hspace{-0.2cm}
		\caption{EVCS cost, User cost and Load variation over each problem set} \label{pf_summarization}
	\end{figure}

	The TOU tariff is changed according to the request of charging/discharging activities and the base load for the EVCS operation. For the users, the best time period to perform the charging is the off-peak hour when there are less requests. Similarly, it is better to perform the discharging during peak-hours, where the TOU tariff is quite high. However, this decreases the benefit of the EVCS, and thus the solution over Eq.~\ref{eq:EVC_obj} prevents the charging and discharging schedule that can lead to the minimal cost of users and vice versa. Also, the network impact objective (Eq.~\ref{obj:3}) leads to the flattened load over each time period, which indeed results in reducing the profit of the EVCS. By balancing the benefits of users, the EVCS and network impact, the charging/discharging schedule obtained by MOEVCS results in a set of good solutions in the PFs so that the stakeholders in the distribution network can choose the solution that satisfies their requirement. 
	
		\begin{table}[!h]
		\centering
		\caption{The range of the objective function values on the PFs}
		\footnotesize
		\label{results_range}
		\begin{tabular}{|p{1cm}|ccc|}
			\hline
			Problems & \multicolumn{1}{c|}{\begin{tabular}[c]{@{}c@{}}User cost\\ ($f_1$)\end{tabular}} & \multicolumn{1}{c|}{\begin{tabular}[c]{@{}c@{}}EVCS cost\\ ($f_2$)\end{tabular}} & \multicolumn{1}{c|}{\begin{tabular}[c]{@{}c@{}}Load variation\\ ($f_3$)\end{tabular}} \\ \hline
			Set 1  & {[}1905, 1946{]}  & {[}-725, -684{]}  & {[}249641, 252625{]}   \\ \cline{1-1}
			Set 2  & {[}2864, 2922{]}  & {[}-1431, -1374{]}  & {[}345844, 350112{]}   \\ \cline{1-1}
			Set 3  & {[}2894, 2941{]}  & {[}-1429, -1475{]}  & {[}353632, 356374{]}   \\ \cline{1-1}
			Set 4  & {[}2956, 3010{]}  & {[}-1540, -1488{]}  & {[}351149, 354238{]}   \\ \hline
		\end{tabular}
	\end{table}
	
	\begin{table}[!ht]
		\centering
		\caption{The objective function values over the baselines and the MOEVCS for each problem set}
		\footnotesize
		\label{com_results}
		\begin{tabular}{|c|c|ccc|}
			\hline
			Problems   & Results  & \multicolumn{1}{c|}{\begin{tabular}[c]{@{}c@{}}User cost\\ ($f_1$)\end{tabular}} & \multicolumn{1}{c|}{\begin{tabular}[c]{@{}c@{}}EVCS cost\\ ($f_2$)\end{tabular}} & \begin{tabular}[c]{@{}c@{}}Load variation\\ ($f_3$)\end{tabular} \\ \hline
			\multirow{7}{*}{Set 1} & B1 & 2287 & -942 & 293839   \\
			& B2 & 2394 & -1183   & 300014   \\
			& B3 & 1793 & -573 & 240775   \\
			& B4 & 2254 & -1031   & 271873   \\
			& B5 & 1786 & -577 & 241598   \\
			& MOMinObj2 & 1946 & -725 & 252625   \\
			& MOMinObj1/3 & 1905 & -684 & 249641   \\ \hline
			\multirow{7}{*}{Set 2} & B1 & 3397 & -1761   & 401569   \\
			& B2 & 3835 & -2353   & 413565   \\
			& B3 & 2759 & -1270   & 338724   \\
			& B4 & 3059 & -1565   & 358594   \\
			& B5 & 2748 & -1261   & 338297   \\
			& MOMinObj2 & 2922 & -1431   & 350112   \\
			& MOMinObj1/3 & 2864 & -1374   & 345844   \\ \hline 
			\multirow{7}{*}{Set 3} & B1 & 3615 & -1979   & 423956   \\
			& B2 & 3816 & -2352   & 425536   \\
			& B3 & 2766 & -1302   & 345481   \\
			& B4 & 3164 & -1694   & 368177   \\
			& B5 & 2758 & -1295   & 345736   \\
			& MOMinObj2 & 2941 & -1475   & 356374   \\
			& MOMinObj1/3 & 2894 & -1429   & 353632   \\ \hline
			\multirow{7}{*}{Set 4} & B1 & 3744 & -2108   & 422189   \\
			& B2 & 3926 & -2463   & 428178   \\
			& B3 & 2790 & -1322   & 341631   \\
			& B4 & 3298 & -1826   & 369089   \\
			& B5 & 2763 & -1298   & 340489   \\
			& MOMinObj2 & 3010 & -1540   & 354238   \\
			& MOMinObj1/3 & 2956 & -1488   & 351149   \\ \hline
		\end{tabular}
	\end{table}

	Fig.~\ref{pf_summarization} summarizes the EVCS cost, user cost, and load variation from Fig.~\ref{com_pf} over the designed problem sets. The range of each objective obtained from the MOEVCS is reported in Table.~\ref{results_range}. Since all the values for the EVCS cost are negative, the absolute values of EVCS cost represent a profit. We can observe the range of each objective function over the $ps$ solutions on the PF. The solution on the PF that leads to the minimal user cost and load variation have the least profit for the EVCS. The solution where the EVCS profit reaches the maximum results in the maximal user cost and load variation.

	\subsubsection{Comparison with Baselines} To verify the effectiveness of the MOEVCS, the results over objective functions are compared with the baselines. We choose two results from the PF, i.e., MOMinObj2 and MOMinObj1/3, for comparison. The result of the baselines and the MOEVCS over the four problem sets are reported in Table.~\ref{com_results}. We can observe that, as expected, baseline 2 (i.e., B2) can lead to the highest profit for the EVCS on the investigated problem sets, i.e., 1183, 2353, 2352, 2463. However, it has the highest user cost and load variation in comparison to B1, B3, B4, B5, MOMinObj2, and MOMinObj1/3. If only considering the benefit of the EVCS, the user cost will increase and this may cause negative impact on the network due to the highest load variation. B5 (SOGA with $f_1$) leads to the lowest user cost on all problem sets, i.e., 1786, 2748, 2758, and 2763, but it has the lowest profit (573, 1270, 1302, and 1322) for the EVCS. Similarly, B3 (SOGA with $f_3$) leads to the lowest network impact on all problems sets, but has the lower profit for the EVCS. Only considering one of the objective functions influences the benefit of its competing objective. 
	
	MOEVCS provides the trade-off solutions that can balance the benefits of users/network and profit of the EVCS. From the PFs in Fig.~\ref{com_pf}, the decrease of EVCS cost (increase of benefit) leads to increase of the user cost and load variation. MOMinObj2 has the minimal cost (i.e., maximal profit) and the maximal user cost/load variation in the PF and MOMinObj1/3 has the minimal user cost/load variation in the PF and the maximal EVCS cost (i.e., minimal profit). From Table.~\ref{com_results}, it is clear that in comparison to B1, B2, B3, B4, and B5, MOMinObj2 and MOMinObj1/3 have trade-off result over the investigated objective functions.
	
	\begin{figure}[h!]\centering
		\subfigure[Set 1]
		{\centering\scalebox{0.32}
			{\includegraphics{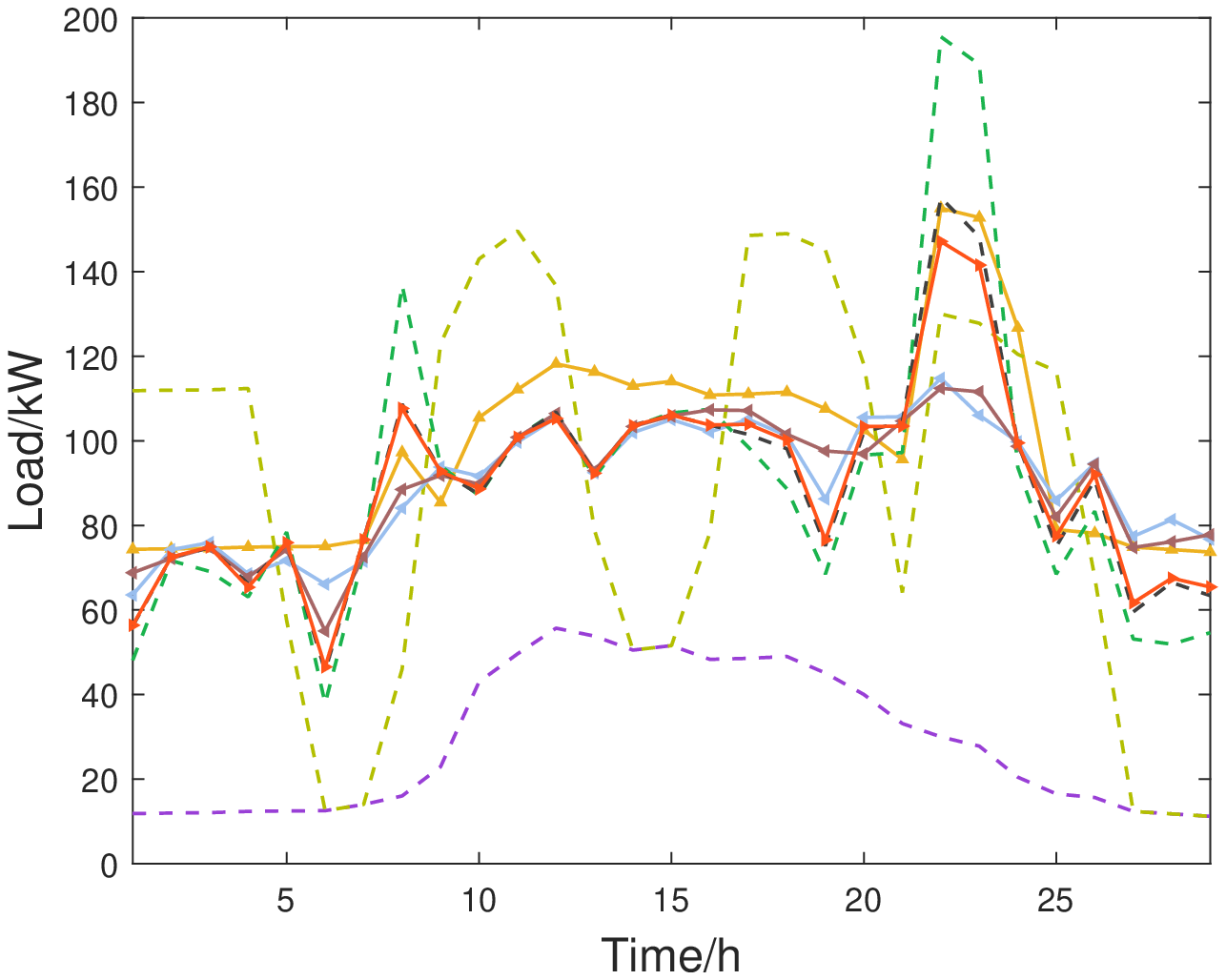}}\label{load_set1}}
		\subfigure[Set 2]
		{\centering\scalebox{0.32}
			{\includegraphics{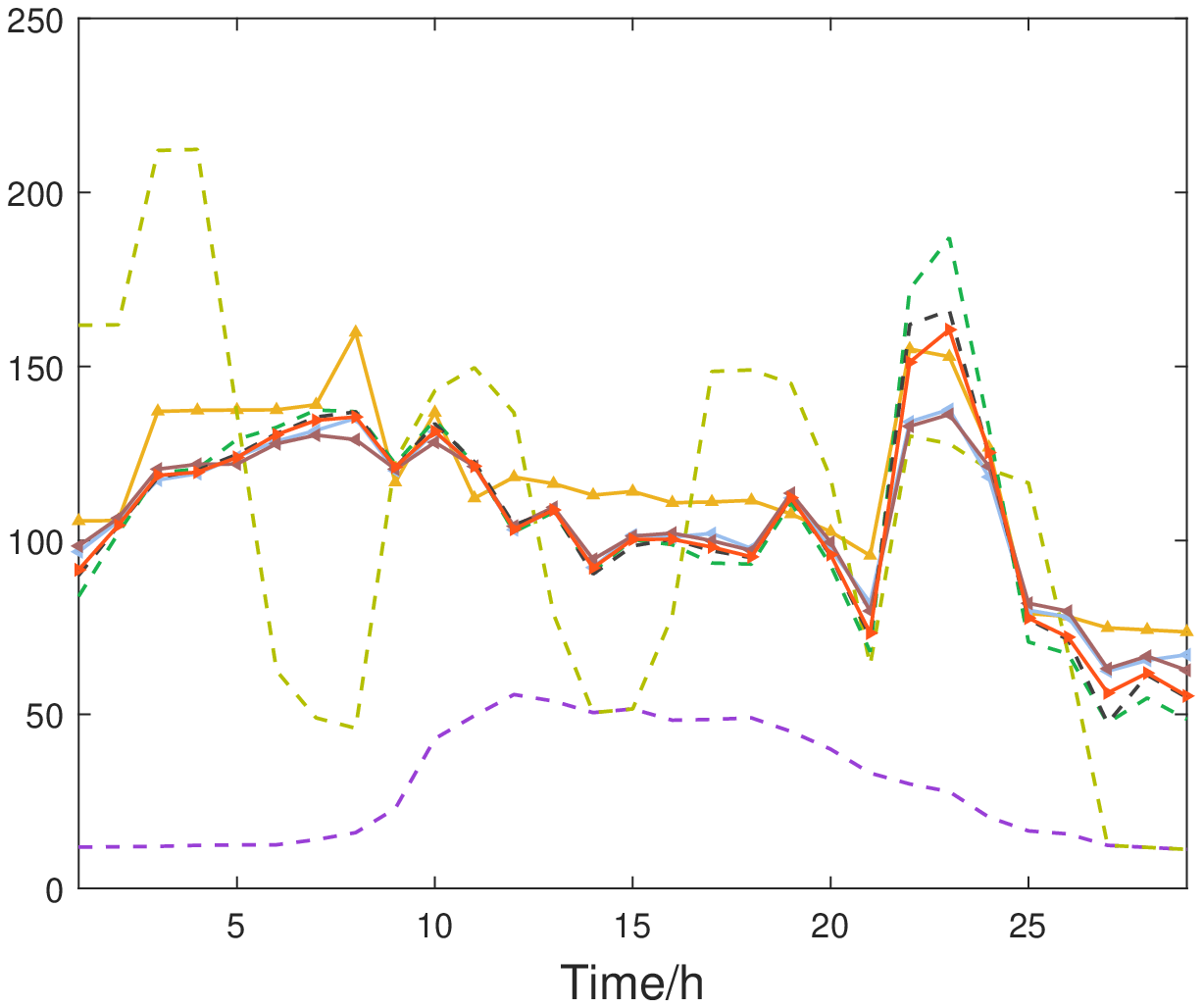}}\label{load_set2}}\\
		\subfigure[Set 3]
		{\centering\scalebox{0.32}
			{\includegraphics{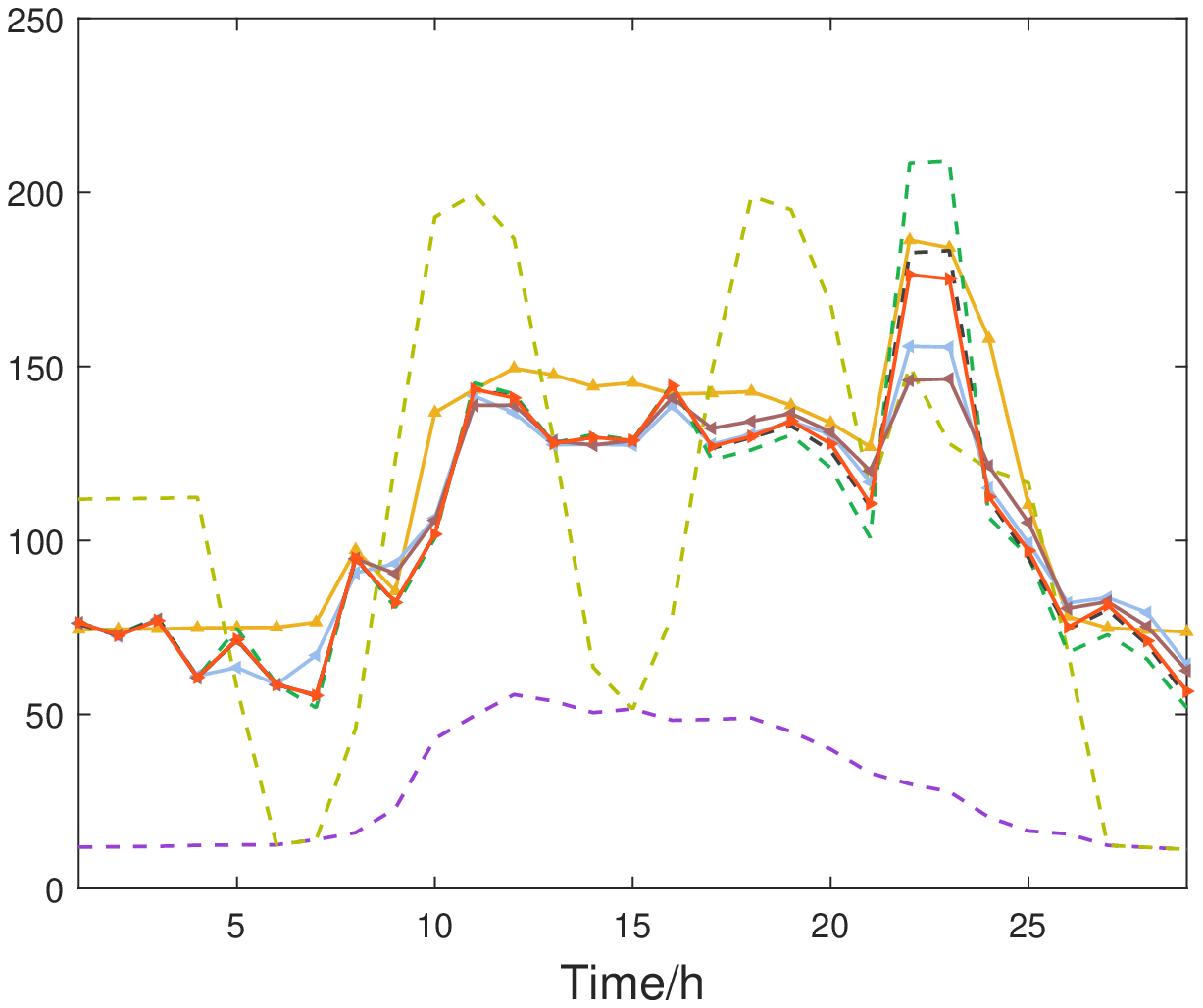}}\label{load_set3}}
		\subfigure[Set 4]
		{\centering\scalebox{0.32}
			{\includegraphics{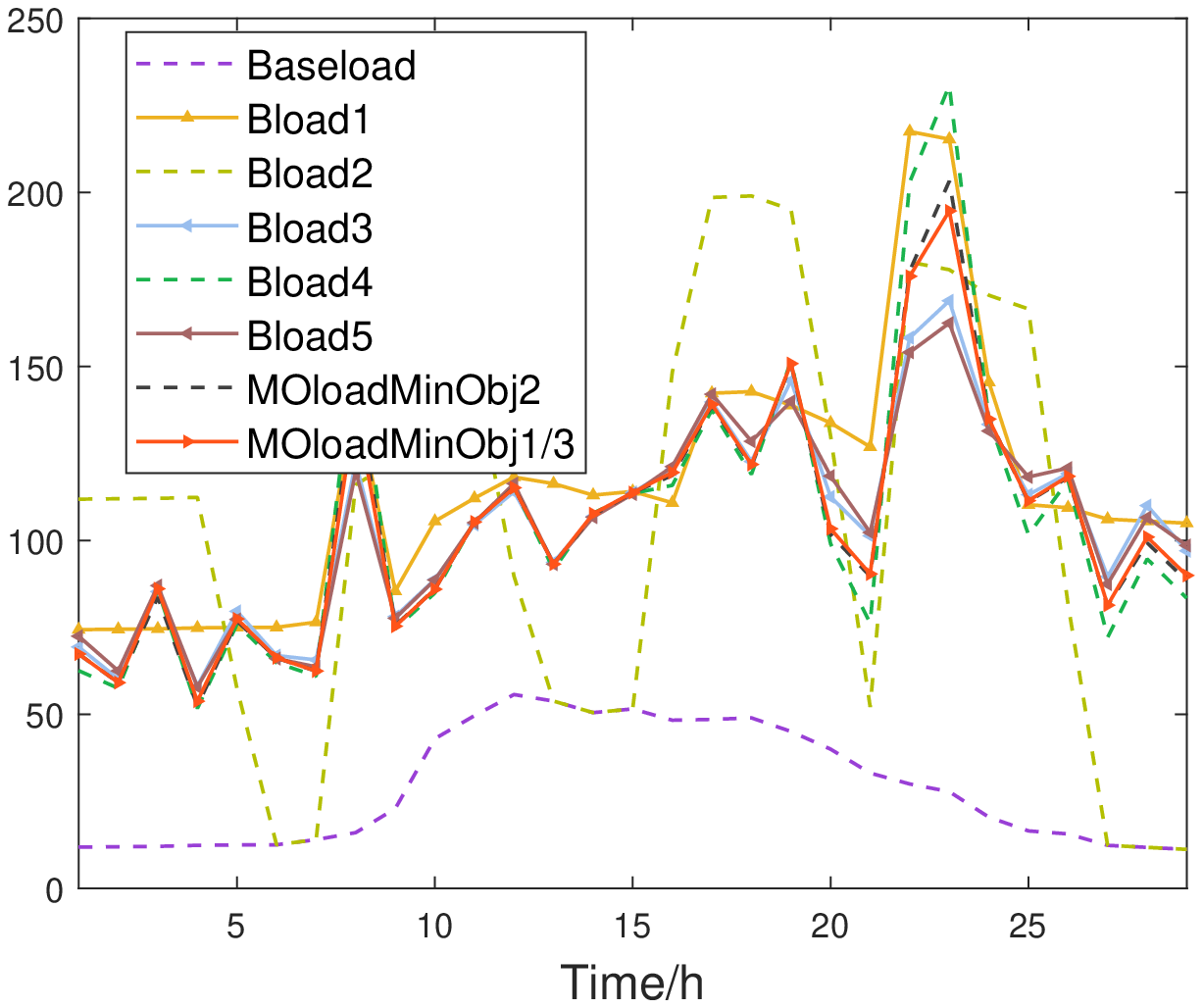}}\label{load_set4}}
		\caption{Load profiling over Bload1, Bload2, Bload3, Bload4, Bload5, MOloadMinObj2 and MOloadMinObj1/3 for each problem set} \label{com_load}
	\end{figure}
	
	\subsubsection{Load Profiling and TOU Tariff} Fig.~\ref{com_load} illustrates the total load (i.e., base load + total charging load + total discharging load) over Bload1, Bload2, Bload3, Bload4, Bload5, MOloadMinObj2 and MOloadMinObj1/3 for each problem set. Bload2 fluctuates significantly and is not good for the distribution network, which is not a good choice for all problem sets. From Table.~\ref{set_distribution}, we can see that the peak hours of the charging request for set 1 are from 22:00 pm to 00:59 am and the Bload1 and Bload4 in Fig.~\ref{load_set1} at 22:00 pm $-$ 23:59 pm are not good for the network in comparison to Bload3, Bload5, and MOloadMinObj1/3. For other problem sets, Bload1 and Bload4 have the similar issue. BLoad3 and BLoad5 show better load profiling performance over the investigated time stamps from the view of load variation and user cost, but they ignore the EVCS cost from the results presented in Table.~\ref{com_results}. Fig.~\ref{load_set2}, Fig.~\ref{load_set3} and Fig.~\ref{load_set4} show the similar load profiling over Bload3, Bload5, MOloadMinObj2 and MOloadMinObj1/3. We also find that the load profiling is coherent with the charging requests distribution in Table.~\ref{set_distribution}.  
	
	\begin{figure}[h!]\centering
		\subfigure[Set 1]
		{\centering\scalebox{0.32}
			{\includegraphics{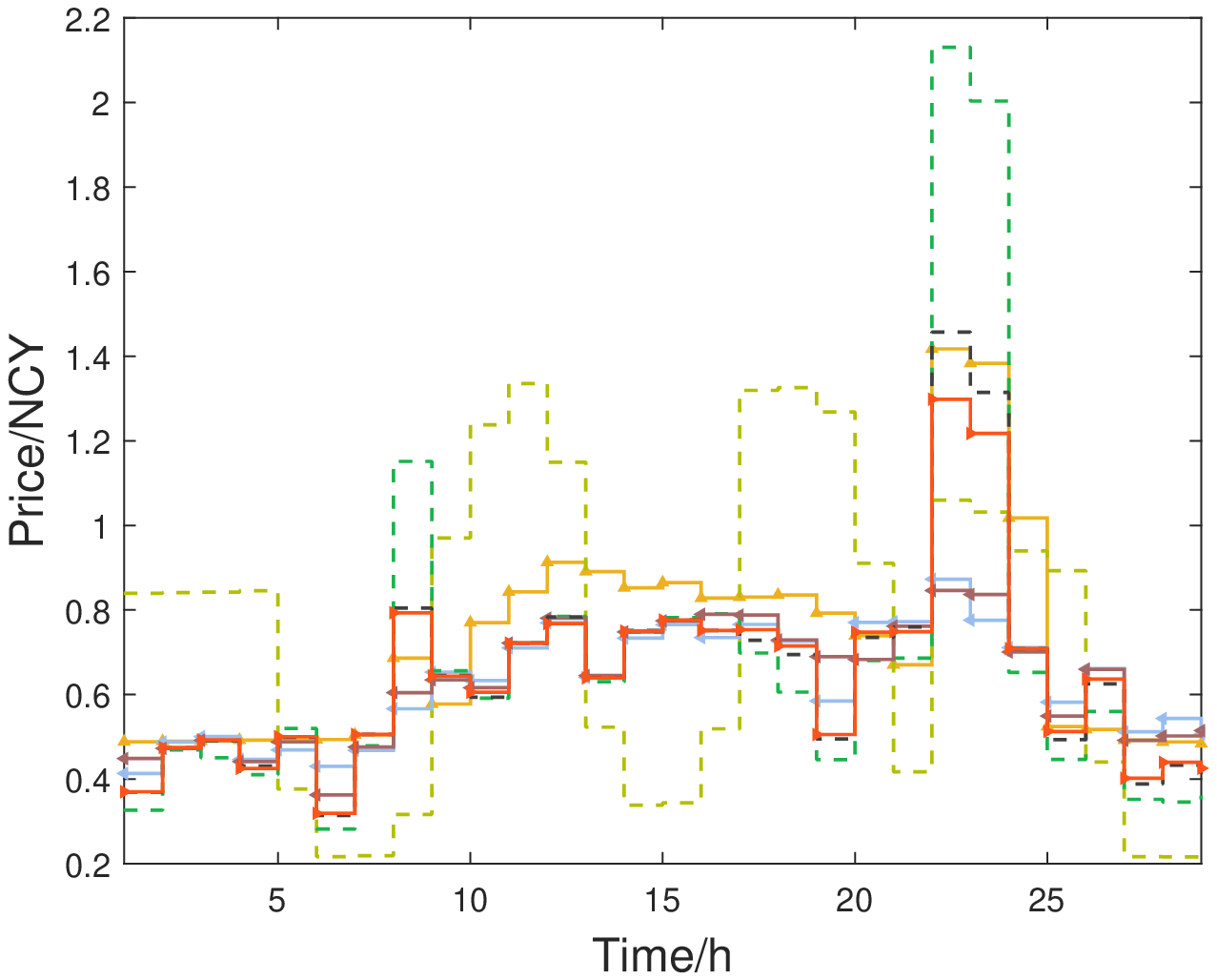}}\label{price_set1}}
		\subfigure[Set 2]
		{\centering\scalebox{0.32}
			{\includegraphics{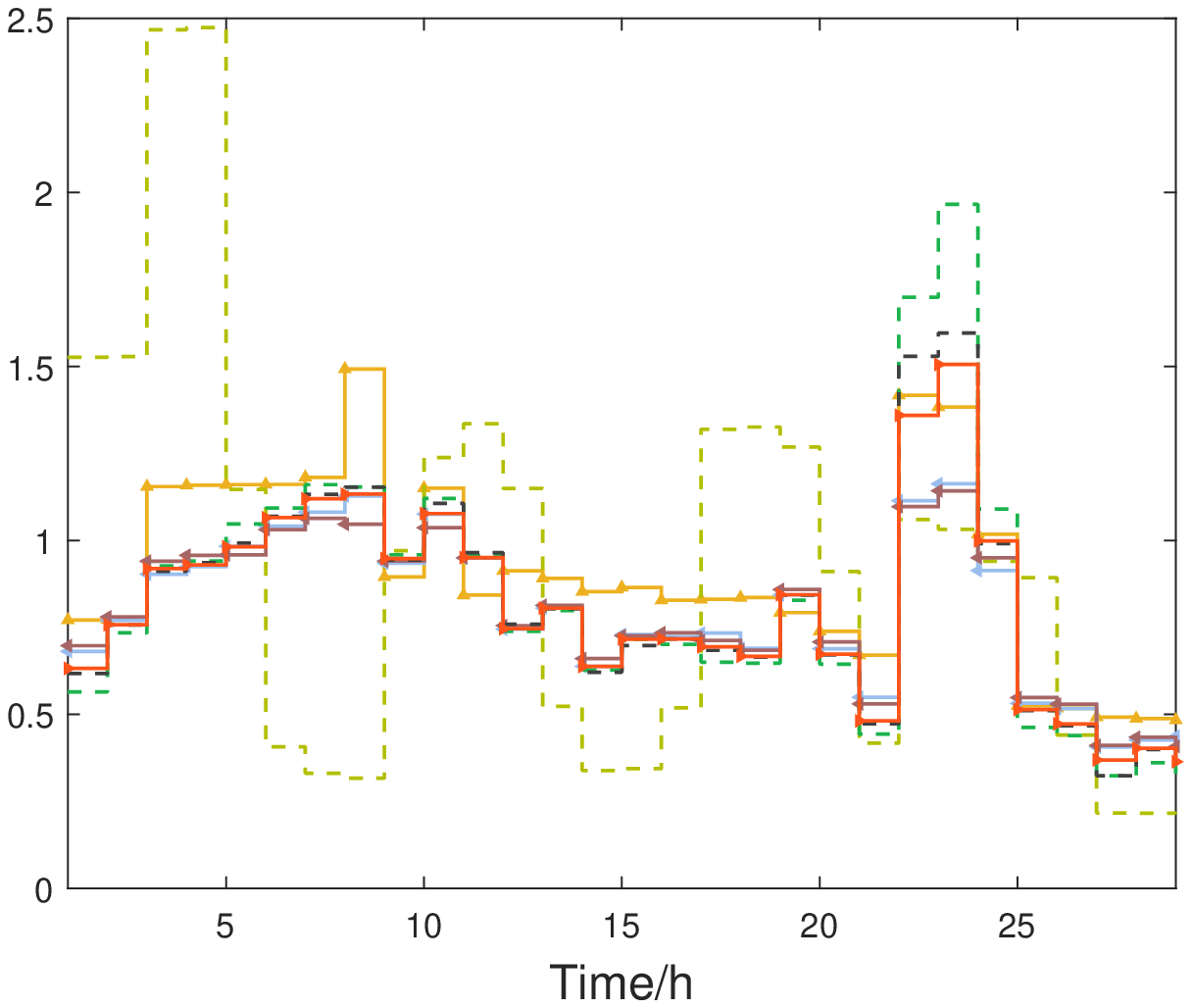}}\label{price_set2}}\\
		\subfigure[Set 3]
		{\centering\scalebox{0.32}
			{\includegraphics{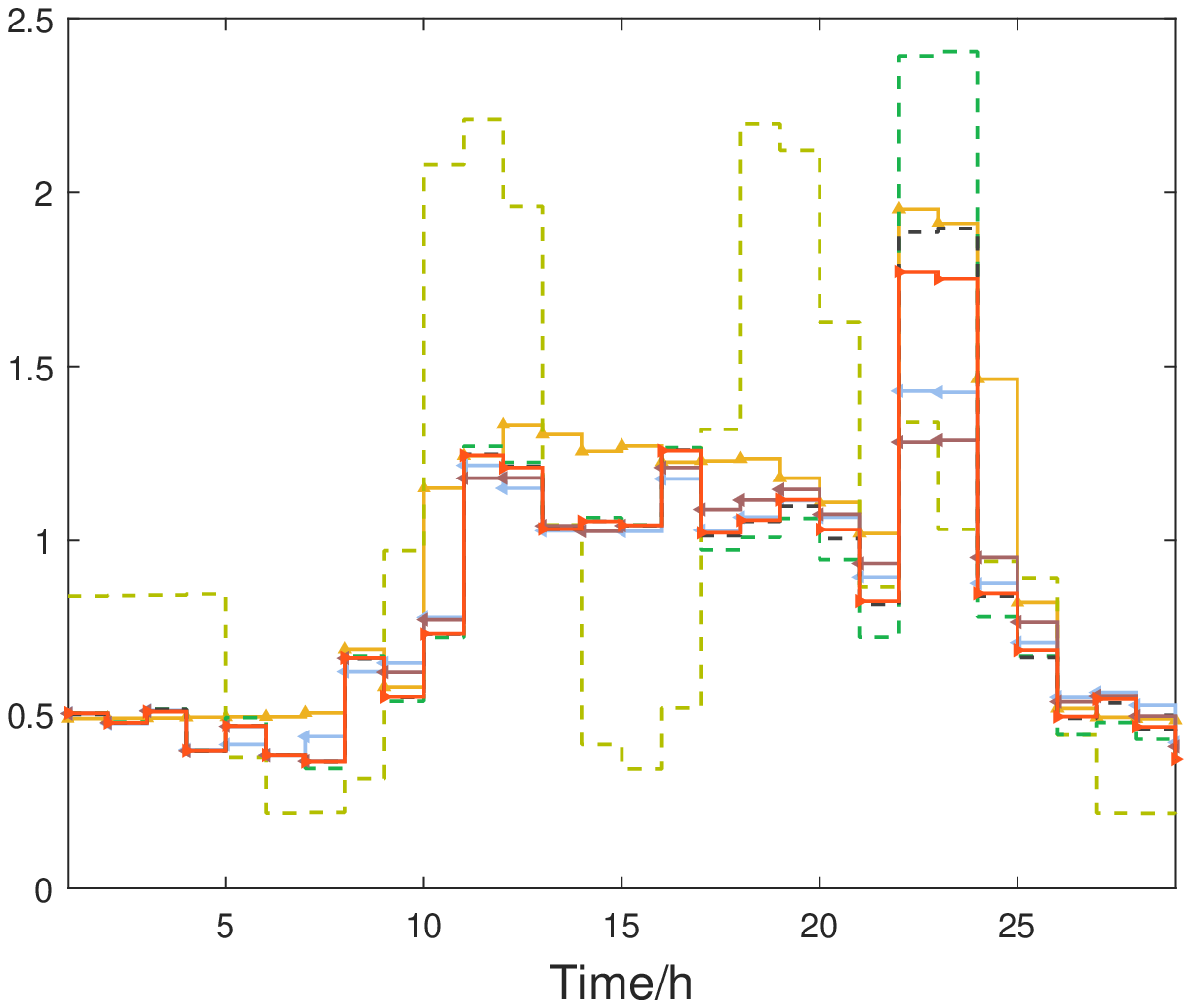}}\label{price_set3}}
		\subfigure[Set 4]
		{\centering\scalebox{0.32}
			{\includegraphics{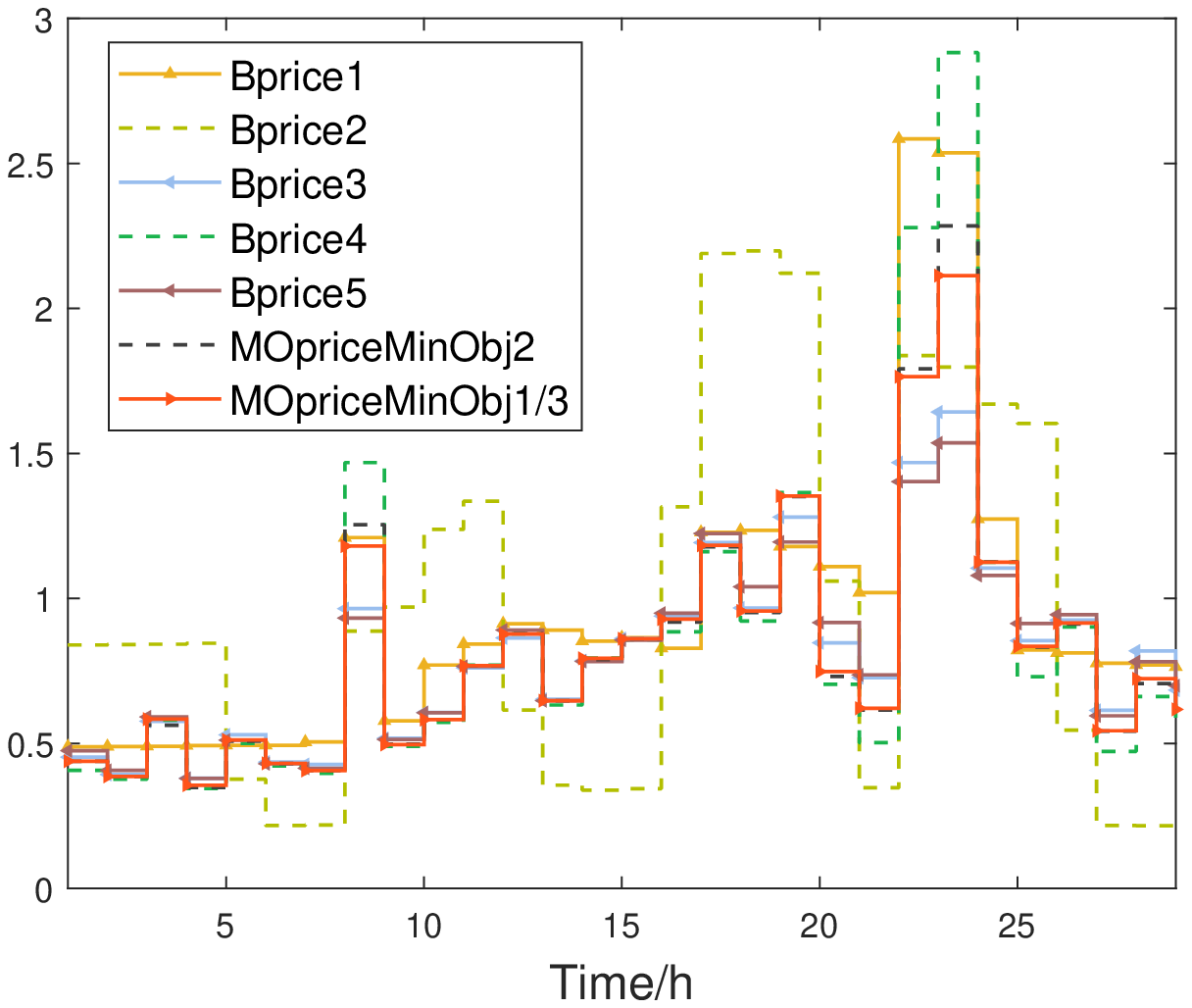}}\label{price_set4}}
		\caption{TOU tariff over Bprice1, Bprice2, Bprice3, Bprice4, Bprice5, MOpriceMinObj2 and MOpriceMinObj1/3 over each problem set} \label{com_price}
	\end{figure}

	Fig.~\ref{com_price} shows the electricity price over different time periods. Since the TOU tariff is decided by Eq.~\ref{eq:pricing}, the TOU tariff is changed according to the total load at each time stamp for each problem set. Comparing Fig.~\ref{com_load} and Fig.~\ref{com_price}, we can see they are quite similar. In this way, optimally scheduling the charging and discharging by considering the TOU tariff help reduce the network impact, given that if more customers choose to charge during the off-peak hours, the off-peak hours will turn to peak hours with the high electricity price.  
	
	\section{Conclusions and Future Works}\label{conclusion}

	In this paper, we proposed a multi-objective electric vehicle charging/discharging schedule (MOEVCS) framework to find the trade-off solutions between three competing objective functions, i.e., the EVCS profit, the user cost and the network impact. We implemented MOEVCS by developing a constraint mixed-variable multi-objective evolutionary algorithm (MVMOEA). To verify the effectiveness of the proposed MOEVCS, we designed five baselines, i.e., without optimization for only charging schedule and single-objective optimization for charging/discharging scheduling with each of the objectives. Time of use (TOU) tariff was designed based on the charging load, discharging load and base load over each time stamp in case of energy spikes, given that most EV owners may choose to charge the EV fleets during off-peak hours and discharge to earn money during peak hours. Instead of focusing on one time period, we designed four different problem sets considering the joint influence of time stamps. In comparison to the baselines, MOEVCS provides the trade-off solutions to balance the benefits from different stakeholders so that they can choose the suitable solution from the PF according to their requirements. In the future, we will focus on investigating public coordinated charging/discharging schedule considering the impact of traffic flow and mobility data.    
	
	\bibliographystyle{IEEEtran}
	\bibliography{References}
	
\end{document}